\documentclass[aps,prd,twocolumn,groupedaddress,showpacs]{revtex4}
\usepackage[dvips]{graphicx}
\usepackage{natbib}

\newcommand{\mboson}{m_{V}}

\newcommand{\vesc}{v_{\mathrm{esc}}}

\newcommand{\dd}{\mathrm{d}}

\newcommand{\erf}{\mathrm{erf}}

\newcommand{\sigx}{\sigma_{\mathrm{xx}}}
\newcommand{\vescr}{v_{\mathrm{esc}}(r)}
\newcommand{\Rsun}{\mathrm{R}_{\odot}}
\newcommand{\Msun}{\mathrm{M}_{\odot}}

\newcommand{\rhox}{\rho_{\mathrm{x}}}
\newcommand{\vbar}{\bar{v}}
\newcommand{\vsun}{v_{\odot}}
\newcommand{\nx}{n_{\mathrm{x}}}
\newcommand{\Tcore}{T_{\odot,\mathrm{c}}}
\newcommand{\rhocore}{\rho_{\odot,\mathrm{c}}}

\newcommand{\Ca}{C_{\mathrm{a}}}
\newcommand{\Cs}{C_{\mathrm{s}}}
\newcommand{\Cc}{C_{\mathrm{c}}}
\newcommand{\Rs}{R_{\mathrm{s}}}

\newcommand{\sigpsi}{\sigma^{\mathrm{SI}}_{\mathrm{p}}}
\newcommand{\sigpsd}{\sigma^{\mathrm{SD}}_{\mathrm{p}}}
\newcommand{\siga}{\sigma_{\mathrm{A}}}
\newcommand{\sigv}{\langle \sigma_{\mathrm{A}} v\rangle}
\newcommand{\Mx}{M_{\mathrm{x}}}
\newcommand{\Nx}{N_{\mathrm{x}}}
\newcommand{\Nxs}{N_{\mathrm{x,s}}}
\newcommand{\Neq}{N_{\mathrm{x,eq}}}
\newcommand{\Neqs}{N_{\mathrm{x,eq}}^{\mathrm{self}}}
\newcommand{\mn}{M_{\mathrm{N}}}
\newcommand{\sign}{\sigma_{\mathrm{N}}}
\newcommand{\fn}{f_{\mathrm{N}}}
\newcommand{\mproton}{m_{\mathrm{proton}}}
\newcommand{\Eth}{E_{\mathrm{TH}}}
\newcommand{\xv}{x_{\mathrm{v}}}

\newcommand{\cmsq}{\ ~\mathrm{cm}^2}

\newcommand{\kms}{\ ~\mathrm{km s}^{-1}}

\begin{document}

\title{
High-Energy Neutrinos From Dark Matter Particle Self-Capture Within the Sun
}
\author{Andrew R. Zentner}
\affiliation{
Department of Physics and Astronomy, University of Pittsburgh, Pittsburgh, PA 15260, USA
}

\date{\today}

\begin{abstract}

A potential flux of high-energy neutrinos from 
the annihilation of dark matter particles trapped within the Sun 
has been exploited to place indirect limits on particle dark matter.  In most models, the dark matter 
interacts weakly, but the possibility of a dark matter particle with a large cross section for 
elastic scattering on other dark matter particles has been proposed in several contexts.  
I study the consequences of such dark matter self-interactions for the high-energy 
neutrino flux from annihilation within the Sun.  The self-interaction among dark matter particles may allow 
dark matter in the halo to be captured within the Sun by scattering off of dark matter particles that have 
already been captured within the Sun.  This effect is not negligible 
in acceptable and accessible regions of parameter space.  Enhancements in the predicted 
high-energy neutrino flux from the Sun of tens to hundreds of percent can be realized in broad 
regions of parameter space.  Enhancements as large as factors of several hundred may be realized 
in extreme regions of the viable parameter space.  Large enhancements require the dark matter 
annihilation cross section to be relatively small, $\sigv \lesssim 10^{-27} \mathrm{cm}^3\mathrm{s}^{-1}$.  
This phenomenology is interesting.  First, self-capture is negligible for the Earth, 
so dark matter self-interactions break the correspondence between the solar and 
terrestrial neutrino signals.  Likewise, the correspondence between indirect and direct detection 
limits on scattering cross sections on nuclei is broken by the self-interaction.  These broken 
correspondences may evince strong dark matter self-interactions.  In some cases, self-capture 
can lead to observable indirect signals in regions of parameter space where limits from 
direct detection experiments would indicate that no such signal should be observable.

\end{abstract}

\pacs{95.35.+d,95.30.Cq,95.55.Vj,98.35.Gi,98.80.Cq}

\maketitle

\section{Introduction}
\label{section:introduction}
%
%

A great deal of observational evidence indicates that a form of 
non-relativistic, non-baryonic matter constitutes the vast majority 
of mass in the Universe.  The unknown nature of the {\em dark matter} that 
binds galaxies and drives cosmic structure formation remains an important 
problem in cosmology and particle physics.  Among dark matter candidates, weakly-interacting 
massive particles (WIMPs), including the lightest superpartners in supersymmetric 
theories, have received the most attention (for a review, see Ref.~\cite{jungman_etal96}).  
In this paper, I study a potential enhancement in high-energy neutrino fluxes 
from dark matter annihilations within the Sun in models where a WIMP-like dark 
matter particle exhibits relatively strong interactions with itself.

Indirect, astrophysical probes of dark matter are an important 
element of any comprehensive program to identify the dark matter 
unambiguously.  One indirect probe of WIMP dark matter is a 
potentially-detectable flux of high-energy muon neutrinos 
arising from the annihilation of dark matter particles 
captured within the Sun \cite{spergel_press85,press_spergel85,krauss_etal85,silk_etal85,
krauss_etal86,gaisser_etal86,griest_seckel87,srednicki_etal87,gould87a,gould87b,gould92}.  
A similar signal from within the Earth may also be exploited in this regard 
\cite{freese86,krauss_etal86,gaisser_etal86}, and though the terrestrial 
signal is typically smaller than the solar signal, it is a valuable cross-check \cite{gould92}.  
In fact, these signals have already been brought to 
bear to limit dark matter elastic scattering cross sections with nucleons 
at interesting levels \cite{ambrosio_etal99,desai_etal04,ackermann_etal06,abbasi_etal09,braun_etal09}.

This basic scenario is simple.  As the Sun moves through the halo of WIMPs, some of the WIMPs 
scatter elastically off of nuclei within the Sun.  Many WIMP-nucleus interactions result in 
WIMPs moving at speeds lower than the local escape speed relative to the Sun.  These particles 
are captured and for a large region of relevant parameter space they come to thermal equilibrium 
in the interior of the Sun.  Eventually, the build-up of WIMPs within the Sun is limited by 
the annihilation of these WIMPs producing neutrinos that can escape from the Sun.  Annihilation 
products other than neutrinos interact within the Sun and are not observable at Earth.

Dark matter particles that interact weakly with standard model particles, 
but exhibit comparably rather strong interactions among themselves have now 
been proposed in several different contexts 
\cite{carlson_etal92,deLaix_etal95,atrio-barandela_davidson97,spergel_steinhardt00,hogan_dalcanton00,mohapatra_teplitz00,dave_etal01,hisano_etal04,hisano_etal05,pospelov_etal08,arkani-hamed_etal08a,lattanzi_silk08,ackerman_etal09,feng_etal09}.  
Some bounds on dark matter self interactions exist 
\cite{yoshida_etal00,gnedin_ostriker01,miralda-escude02,randall_etal08,kamionkowski_profumo08} 
and observational tests of various scenarios have been proposed 
\cite{robertson_zentner09,pieri_etal09,spolyar_etal09,finkbeiner_etal09,slatyer_etal09}, 
but a wide range of parameter space remains and will remain viable for  
$\Mx \sim$ a few $\times 10^2$~GeV dark matter particles with large 
self-interaction cross sections, $\sigx \sim 10^{-24} \cmsq$.

Large cross sections for dark matter particles to scatter elastically off of each other 
open a new possibility for capture within the Sun.    In addition to nuclei, 
previously-captured dark matter particles or dark matter particles otherwise sequestered 
within the solar interior may serve as additional targets for the capture of halo dark matter particles.  
I refer to this as dark matter ``self-capture'' and 
it is this possibility that I consider in detail in this paper.  
Previous studies have considered distinct modifications to high-energy neutrino 
fluxes from the Earth and Sun due to inelastic scattering of dark matter 
with against nuclei \cite{nussinov_etal09,menon_etal09}.

I begin in \S~\ref{section:nus} with a brief sketch of the standard scenario for indirect detection 
of dark matter via high-energy neutrinos from the solar interior.  
In \S~\ref{section:sc}, I use a simple, order-of-magnitude estimate to 
show that dark matter self-capture within the Sun may not be 
negligible for a range of viable and interesting models.  On the other hand, 
self-capture within the Earth is always negligible for models that have not yet 
been excluded by other means.

In \S~\ref{section:results}, I give the results of more detailed calculations of the importance of 
dark matter particle self-capture within the Sun.  These 
results demonstrate that modest enhancements of tens to a 
hundred percent relative to models in which self-capture is negligible are possible over reasonably 
broad ranges of interesting parameter space.  Significantly larger flux enhancements of up to factors of 
hundreds are possible in extreme regions of the dark matter parameter space.  Throughout, I remain 
relatively agnostic about the nature of the dark matter and present results as a function of the most 
directly relevant model parameters, dark matter particle mass $\Mx$, dark matter-proton scattering cross 
section $\sigma_{\mathrm{p}}$, dark matter self-interaction cross section $\sigx$, and thermally-averaged 
dark matter annihilation cross section multiplied by relative velocity $\sigv$.  However, I do use 
the findings of detailed explorations of the parameters available to neutralino dark matter in 
supersymmetric scenarios as guidance for interesting values of these parameters  \cite{baltz_gondolo04,
roszkowski_etal07,baltz_etal08}.

I summarize my results and conclusions in \S~\ref{section:discussion}.  In particular, I emphasize 
that flux enhancements from self-capture scenarios may be important for two reasons.  First, the 
solar flux may be significantly altered by self-capture, while the terrestrial flux cannot be.  Therefore, 
the ratio of the solar to terrestrial high-energy neutrino fluxes from dark matter may be markedly different 
from the standard predictions and this may signify new dark 
matter interactions.  Likewise, the similar correspondence 
between direct search signals and solar high-energy neutrino fluxes can be broken.  In some cases, models that 
may otherwise be ruled out by direct dark matter searches may produce observable neutrino signals due to the 
self-capture enhancement.  I include in an Appendix the details of the capture rate calculations that I perform, 
including an example of the capture rates that I compute in 
the standard scenario of spin-independent capture off 
of nuclei.  This discussion follows the derivations given by Gould \cite{gould87b,gould92}.

\section{High-Energy Neutrinos from the Sun and Dark Matter Self-Capture}
\label{section:nus}

In the most well-studied scenarios, captured dark matter particles typically thermalize in 
the solar interior on a timescale less than the age of the Sun 
($\tau_{\odot} \approx 5 \times 10^9$~yr) as well as the other timescales in the problem 
\cite{spergel_press85,krauss_etal85,silk_etal85,krauss_etal86,gaisser_etal86,griest_seckel87,srednicki_etal87,gould87b}.  
In this case, the time evolution of the number of dark matter particles in the Sun 
$\Nx$, follows 
\begin{equation}
\label{eq:riccati}
\frac {\dd \Nx} {\dd t} = \Cc + \Cs \Nx - \Ca \Nx^2.
\end{equation}
The coefficients are the rate of capture of dark matter particles by scattering off 
of nuclei within the Sun $\Cc$, {\em twice} the rate of annihilation per pair of 
dark matter particles within the Sun $\Ca$ 
({\em twice} because each annihilation eliminates two particles), 
and the rate of capture of dark matter particles by scattering off of 
other dark matter particles that have already been captured within the Sun $\Cs$.  The $\Cs \Nx$ 
term on the right-hand side of Eq.~(\ref{eq:riccati}) is the new term that I study in 
this paper.  Capture rates were first computed  by 
\citet{press_spergel85} and this calculation was revised, corrected, and 
greatly expanded upon in an impressive series of papers by Gould \cite{gould87a,gould87b,gould92}.  
In principle, evaporation of captured particles from the Sun can also occur, but this 
is unimportant for masses larger than a few GeV \cite{gould87a,griest_seckel87}.  I discuss 
the specific rates at greater length below and in the appendix.  For the time being, let 
us focus attention on Eq.~(\ref{eq:riccati}).

In the standard treatment, self-capture of dark matter particles is ignored ($\Cs=0$).  The solution 
of Eq.~(\ref{eq:riccati}) for $\Nx=0$ at $t=0$ is then
\begin{equation}
\label{eq:rsol1}
\Nx = \sqrt{\frac{\Cc}{\Ca}} \tanh ( \sqrt{\Cc \Ca} t).
\end{equation}
There is a timescale for equilibration between dark matter annihilation and dark matter capture, 
$\tau_{\mathrm{eq}} = 1/\sqrt{\Cc \Ca}$.   For many models of interest, $\tau_{\mathrm{eq}} << \tau_{\odot}$ 
and the solution approaches a steady state, 
\begin{equation}
\label{eq:rsteady1}
\Neq = \sqrt{\frac{\Cc}{\Ca}}.
\end{equation}
In this circumstance, the rate of annihilation of captured dark matter particles within 
the Sun is  
\begin{equation}
\label{eq:gsteady1}
\Gamma_{\mathrm{a}} = \frac{1}{2} \Ca (\Neq)^2 = \frac{1}{2}\Cc.
\end{equation}
The factor of $1/2$ in Eq.~(\ref{eq:gsteady1}) arises because there are 
$\Neq^2/2$ distinct pairs of particles, while this factor is not present 
in Eq.~(\ref{eq:riccati}) because each annihilation eliminates two dark matter 
particles from the Sun.  
The annihilation rate $\Gamma_{\mathrm{a}}$ is independent of the annihilation rate 
coefficient $\Ca$.  Consequently, the observable flux at Earth is independent of 
the dark matter mutual annihilation cross section, provided the cross section is 
large enough that the equilibrium solution [Eq.~(\ref{eq:rsteady1})] obtains.  
Dark matter particles annihilate upon capture and the flux 
at the Earth is modulated only by the capture rate $\Cc$.

In the first papers to study high-energy neutrinos from dark matter annihilation 
within the Sun, it was noted that this phenomenology is interesting and useful.  First, 
the flux from annihilations at rate Eq.~(\ref{eq:gsteady1}) is independent of annihilation 
cross section, so this is an indirect search method that does not rely on models with 
relatively large annihilation cross sections.  Second, the high-energy neutrino 
signal from annihilation in the solar interior can be related to other potentially-observable signals.  
I discuss the rate of capture in the 
appendix, but it suffices to note that the capture rate of dark matter particles in the Sun should 
be proportional to the cross section for a dark matter particle to scatter off of a 
nucleus ($\sign$) within the Sun and the local density of dark matter ($\rhox$), 
$\Cc \propto \sign \rhox$.  A similar process may operate within the Earth, whereby captured 
dark matter particles give rise to high-energy neutrinos from 
the Earth's center and this signal should 
also grow in proportion to the product $\sign \rhox$ \cite{freese86,gaisser_etal86,krauss_etal86}.  
Moreover, signals in direct dark matter search experiments 
are proportional to this same product as well \cite{jungman_etal96}.  As a consequence, 
indirect detection of dark matter through high-energy neutrinos from 
the Sun and Earth as well as direct dark matter search 
experiments may serve to check each other and corroborate any detections.

\begin{figure}[t]
\includegraphics[height=7cm]{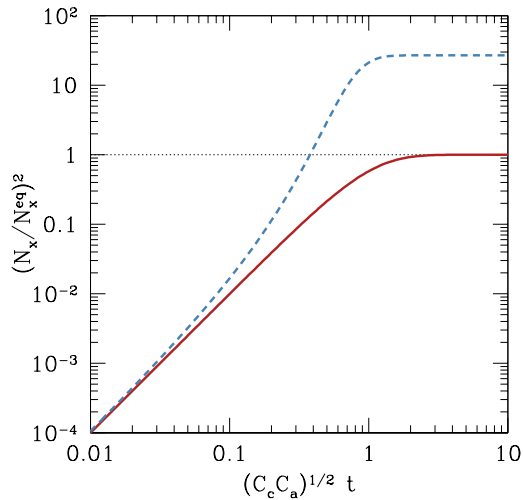}
\caption{
The evolution of the annihilation rate of 
dark matter particles captured within the sun as a function 
of time.  The {\em solid} lines shows evolution to the 
steady-state solution of Eq.~(\ref{eq:rsteady1}) and Eq.~(\ref{eq:gsteady1}) 
in the absence of significant dark matter self interactions.  
The {\em dashed} lines show evolution to the new equilibrium in 
a model in which self-interactions are important and 
$\Cs = 5\sqrt{\Cc \Ca}$.  Time is shown in units of the standard 
equilibration timescale $\tau_{\mathrm{eq}}=1/\sqrt{\Cc \Ca}$ and 
the squared number of captured dark matter particles is shown 
relative to the standard equilibrium number $\Neq=\sqrt{\Cc/\Ca}$.
}
\label{fig:nuex}
\end{figure}

If self-capture is not ignored this picture may be altered.  In particular, the general solution 
to Eq.~(\ref{eq:riccati}) with $\Cs \ge 0$ is 
\begin{equation}
\label{eq:rsol2}
\Nxs = \frac{ \Cc \tanh (t/\zeta)}
{\zeta^{-1} - \Cs \tanh( t/\zeta)/2}
\end{equation}
where
\begin{equation}
\label{eq:etadef}
\zeta = \frac{1}{\sqrt{\Cc \Ca + \Cs^2/4}}.
\end{equation}
The steady-state solution at $t \gg \zeta$ is 
\begin{equation}
\label{eq:rsteady2}
\Neqs = \frac{\Cs}{2\Ca} + \sqrt{\frac{\Cs^2}{4\Ca^2} + \frac{\Cc}{\Ca}}.
\end{equation}
Clearly, Eq.~(\ref{eq:rsteady2}) reduces to Eq.~(\ref{eq:rsteady1}) in the 
case that $\Cs=0$.  Whether or not self-capture of dark matter particles 
is important can be discerned by comparing the timescales $\Cs^{-1}$ and 
$1/\sqrt{\Cc \Ca}$, and I will pursue this comparison shortly.  Consider the 
case where $\Cs^2 \gg \Cc \Ca$.  In this circumstance, $\Neqs \simeq \Cs/\Ca$.  
The annihilation rate within the sun is then 
\begin{equation}
\label{eq:gsteady2}
\Gamma_{\mathrm{a}} = \frac{1}{2}\Ca (\Neqs)^2 = \frac{1}{2} \Cs^2/\Ca.
\end{equation}
The annihilation rate grows in inverse proportion to the rate coefficient 
$\Ca$ when self-capture is possible.  As I discuss below, the annihilation 
rate coefficient is proportional to the annihilation cross section, so models 
with relatively {\em low} annihilation cross sections are favored for an 
indirect neutrino signal from the Sun when self-capture is not negligible.  

Figure~\ref{fig:nuex} shows two examples of the evolution of the rate of annihilation 
of dark matter particles captured within the Sun as a function of time.  If self-capture 
is not negligible, the evolution of the total number of captured dark matter particles 
may come to be dominated by the second term on the right-hand side of Eq.~(\ref{eq:riccati}) 
and $\Nx$ may grow exponentially for some time.  This exponential growth is eventually terminated by 
annihilations, so the lower the annihilation cross section, the more exponential growth 
of $\Nx$ is important and the greater the relative flux of neutrinos from the Sun 
may be (provided the Sun is old enough that the equilibrium solution has been achieved).

\section{The Importance of Dark Matter Self-Capture}
\label{section:sc}

I now present a simple estimate of the parameter values for which dark matter particle self-capture 
may be a non-negligible effect.  I define the ratio $\Rs = \Cs^2/\Cc \Ca$.  Comparing to 
Eq.~(\ref{eq:rsol2}) and Eq.~(\ref{eq:rsteady2}), self-capture will be negligible when 
$\Rs \ll 1$ and dominant when $\Rs \gg 1$.  The task now is to evaluate each of the 
rate coefficients for a set of parameters describing the interactions of the dark matter 
particle.

I show in the Appendix that a simple approximation for the self-capture rate is 
\begin{equation}
\label{eq:Cselftext}
\Cs \Nx \approx \sqrt{\frac{3}{2}}\ \nx \sigx \vesc(\Rsun)\ \frac{\vesc(\Rsun)}{\vbar}  \Nx 
\langle \hat{\phi}_{\mathrm{x}} \rangle \frac{\erf(\eta)}{\eta},
\end{equation}
where $\nx$ is the local number density of dark matter particles in the halo, 
$\sigx$ is the dark matter elastic scattering cross section, 
$\vesc(\Rsun)$ is the escape speed from the surface of the Sun, 
$\vbar$ is the local three-dimensional velocity dispersion of dark 
matter particles in the halo, 
$\langle \hat{\phi}_{\mathrm{x}} \rangle \simeq 5.1$ is a dimensionless 
average solar potential experienced by captured dark matter particles within 
the Sun, and $\eta^2 = 3(\vsun/\vbar)^2/2$ is the square of a 
dimensionless speed of the Sun through the Galactic halo.  Relation 
Eq.~(\ref{eq:Cselftext}) neglects the recoils of the target dark matter particles; however, I 
demonstrate in the Appendix that this is a reasonable approximation when escape speeds 
are significant compared to $\vsun$ and $\vbar$.  This condition holds for 
capture within the Sun.

The full expressions for capture off of nuclei are unwieldy, so I make an effort here to 
evaluate $\Rs$ approximately.  To simplify the evaluation of $\Rs$, I temporarily 
assume that capture of dark matter by scattering off of nuclei is never significantly 
restricted by the kinematics of the scattering.  This is not generally true.  In any 
individual scattering event dark matter particles can lose a fraction of their kinetic 
energy $\Delta E/E \le 4\Mx \mn/(\Mx + \mn)^2$, where $\mn$ is the mass of the target nucleus.  
Capture within the Sun typically requires $\Delta E/E \gtrsim 1/5$.  Therefore, capture 
can be efficient for dark matter and nucleus masses that differ by a factor of as much 
as $\sim 20$.  If the mass of the dark matter particle is sufficiently different from 
the mass of the nucleus on which it scatters, capture may be kinematically unfavorable.

Neglecting kinematic limitations to capture on nuclei, the capture rate 
off of a particular nuclear species $N$ is given by a 
relation analogous to Eq.~(\ref{eq:Cselftext}) with $\Cs \Nx \rightarrow \Cc$, 
$\sigx \rightarrow \sign $, $\Nx \rightarrow \fn M_{\odot}/\mn$, and 
$\langle \hat{\phi}_{\mathrm{x}} \rangle \rightarrow \langle \hat{\phi}_{\mathrm{N}} \rangle$.  
The quantity $\sign$ is the scattering cross section of the dark matter particle off 
of the nucleus $N$, $\fn$ is the fraction of the solar mass in nucleus $N$, 
and $\langle \hat{\phi}_{\mathrm{N}} \rangle$ is the average dimensionless potential 
experienced by these nuclei.  For most nuclei within the Sun, 
$\langle \hat{\phi}_{\mathrm{N}} \rangle \simeq 3.2$ \cite{gould92}.  
Examining Eq.~(\ref{eq:Cselftext}), the capture rate off of nuclei in 
this limit scales with dark matter mass as $\Cc \propto \Mx^{-1}$.  
When $\Mx \gg \mn$, the kinematic limitation to the capture rate is not 
negligible, the maximum fractional kinetic energy lost per collision 
is $\sim 4\mn/\Mx$, and the capture rate scales as $\Cc \propto \Mx^{-2}$.

Assuming that the dark matter particles equilibrate with the solar interior 
rapidly upon capture, the coefficient $\Ca$ is likewise simple to estimate.  
Let $\epsilon_{\mathrm{x}}(r)$ be the number density of dark matter particles 
as a function of position in the Sun.  The annihilation rate coefficient is 
then 
\begin{equation}
\label{eq:ca1}
\Ca = \frac{4 \pi \sigv}{\Nx^2} \int_0^{\Rsun}\ \epsilon^2_{\mathrm{x}}(r)\ r^2 \dd r,
\end{equation}
and the naive expectation for a non-relativistic thermal relic 
dark matter particle is that $\sigv \sim 10^{-26}\ \mathrm{cm}^3\mathrm{s}^{-1}$.  
Under the assumption of a thermal distribution at an effective 
solar core temperature $\Tcore=1.57 \times 10^7\ \mathrm{K}$, the distribution 
$\epsilon_{\mathrm{x}}(r) \propto \exp[-\Mx \phi(r)/\Tcore$], where $\phi(r)$ is 
the gravitational potential as a function of position within the Sun.  
Making a further assumption of a constant solar density of 
$\rhocore = 150\ \mathrm{g}/\mathrm{cm}^3$ the integral is straightforward 
to evaluate.  Conventionally, this has been represented in terms of effective volumes 
\begin{equation}
\label{eq:ca2}
\Ca = \sigv \frac{V_2}{V_1^2}
\end{equation}
where
\begin{equation}
\label{eq:veff}
V_{\mathrm{j}} = 2.45 \times 10^{27} \Bigg(\frac{100 \mathrm{GeV}}{\mathrm{j}\Mx}\Bigg)^{3/2} \ \mathrm{cm}^3.
\end{equation}
The effective volumes represent $\sim 10^{-6}$ of the total solar volume, so captured particles 
extend over only $\sim 1\%$ of the solar radius, 
justifying the constant density approximation \cite{griest_seckel87}.

The question arises whether the strong self-interactions among the dark matter 
particles should alter the assumption of a thermal distribution.  A definitive 
answer to this question requires solving the Boltzmann equation.  Such a 
calculation is extensive and beyond the scope of this paper.   
It is reasonable to suspect that any modifications will be minor 
for parameters of interest.  In the relevant parameter regime 
(which will be more clearly delineated below) the 
increase in the number of captured dark matter particles relative to the 
standard scenario is modest (less than a factor of $\sim 30$), yet 
collisions among dark matter particles may happen at a rate that is comparable 
to the rate of collisions between dark matter particles and nuclei.  
It is a relatively simple matter to use the approximate methods of 
Ref.~\cite{gould87b} to show that particles with $\Mx \gtrsim 10$~GeV 
remain localized well within the solar interior and are not altered by 
the temperature gradient within the Sun, and that the rate of 
energy inflow due to capture is slower than the rate of thermalization 
with the solar interior.  Finally, it is also straightforward to show 
that for masses $\lesssim 10^{12}$~GeV 
(which far exceeds the unitarity bound for a thermal relic, 
e.g. \cite{griest_kamionkowski90}), the dark matter density is never 
sufficiently large for the dark matter to be self-gravitating, so the 
self-interaction does not lead to rapid collapse via the gravothermal 
catastrophe (the case of $\Mx \gtrsim 10^{12}$~GeV is treated in 
Ref.~\cite{albuquerque_etal01}).  
Given these considerations, a standard thermal profile seems a 
reasonable approximation and I proceed under this assumption.

\begin{widetext}
With expressions for $\Cc$, $\Cs$, [Eq.~(\ref{eq:Cselftext})] 
and $\Ca$ [Eq.~(\ref{eq:ca2})], all of the pieces are now in place 
to approximate the ratio $\Rs = \Cs^2/\Cc \Ca$.  Making the simplifying 
assumption that capture occurs primarily off of a single type of nucleus, 
this gives 
\begin{equation}
\label{eq:rssym}
\Rs \approx \sqrt{\frac{3}{2}} \frac{\sigx^2 \vesc(\Rsun)}{\sign \sigv}\ \frac{\vesc(\Rsun)}{\vbar}\ 
\frac{\langle \hat{\phi}_{\mathrm{x}} \rangle^2}{\langle \hat{\phi}_{\mathrm{N}} \rangle}\ \frac{V_1}{V_2}\ 
\frac{\nx\ \mn V_1}{\fn \Msun}\ \frac{\erf(\eta)}{\eta}.
\end{equation}
Evaluating this for the particular case of capture off of Oxygen (which is the most important 
individual element for dark matter capture in the Sun, 
see Figure~\ref{fig:captrate} in the Appendix and Ref.~\cite{gould92}), 
taking $\fn = 10^{-2}$ (almost twice the solar Oxygen abundance to account for the 
simplicity of the current estimate), and keeping relevant aspects of the dark matter 
particle model explicit yields, 
\begin{equation}
\label{eq:rsnum}
\Rs \approx 0.4 \Bigg(\frac{\sigx}{10^{-24} \ \mathrm{cm}^2}\Bigg)^2
\Bigg(\frac{10^{-42}\ \mathrm{cm}^2}{\sign}\Bigg) 
\Bigg(\frac{10^{-27}\ \mathrm{cm}^3\mathrm{s}^{-1}}{\sigv}\Bigg) 
\Bigg(\frac{100\ \mathrm{GeV}}{\Mx}\Bigg)^{5/2}.
\end{equation}
\end{widetext}

The result in Eq.~(\ref{eq:rsnum}) is somewhat startling.  Current best bounds on 
the elastic scattering cross section of dark matter particles off of each other are 
assumption-dependent and approximate, but they indicate that 
$\sigx \lesssim 10^{-23}\ (\Mx/100\ \mathrm{GeV}) \ \mathrm{cm}^2$ 
\cite{yoshida_etal00,dave_etal01,gnedin_ostriker01,miralda-escude02,randall_etal08}.  
A slightly less restrictive bound from an analysis of the bullet cluster 
is probably the least dependent upon particular assumptions \cite{randall_etal08}.  
Eq.~(\ref{eq:rsnum}) indicates that dark matter particle self-capture 
within the Sun is not necessarily a negligible effect in 
acceptable regions of dark matter particle parameter space.  Moreover, 
Eq.~(\ref{eq:rsnum}) neglects the fact that dark matter capture off of nuclei 
may be kinematically unfavorable, while self-capture can never be kinematically 
unfavorable because the particles will always have equal mass.

I note in passing that self-capture by the Earth can never be important.  
As I discuss in the Appendix, collisions between halo dark matter 
particles and particles already captured within a body may result in the target 
dark matter particles being ejected from the body upon recoil.  
The net result in this case is no gain in the number of captured dark 
matter particles.  Ejection by recoil depends upon the ratio of the speed 
of the particles at infinity to the escape speed from the body.  Within the Sun, 
escape speeds are always significantly higher than the typical relative speeds 
of dark matter particles at infinity and ejection is only a small correction to the 
simple solar capture estimate.  In the case of the Earth, escape speeds are more than 
an order of magnitude {\em lower} than the typical relative speeds of dark matter particles at 
inifinity so almost all collisions within the Earth result in ejection of the 
target dark matter particle.  
The ejection of the targets from the Earth introduces 
the possibility that the halo dark matter particles 
may scour the Earth of any particles captured through interactions 
with nuclei.  This can be computed in a manner analogous to self-capture, though 
the sign of the term linear in $\Nx$ in Eq.~(\ref{eq:riccati}) 
would be negative.  In the case of the Earth, the ejection rate 
is small and $\Rs \lesssim 10^{-3}$ for all parameters of interest.  Modifications 
to the Earth signal are negligible.

In the following section, I show results from a more detailed calculation of the importance 
of dark matter particle self-capture for high-energy neutrino fluxes observed at the Earth.  
I use the formulae from Ref.~\cite{gould92} to compute the capture rate of dark matter 
particles from nuclei as described in the Appendix.  These formulae are lengthy and I do not reproduce 
them in full here, though I give an example of the capture rates that I use in Fig.~\ref{fig:captrate}.  
I use Eq.~(\ref{eq:Cselffinal}) to compute the rate coefficient for 
self-capture of dark matter particles.  This relation is derived in the Appendix and 
includes the reduction in the capture rate due to the potential ejection of 
target dark matter particles.  

The most relevant quantity to compute is the 
{\em enhancement} in the neutrino signal due to self-capture of dark matter relative to 
the neutrino flux expected in the standard calculation.  I define the quantity 
\begin{equation}
\label{eq:betadef}
\beta \equiv \frac{\Nxs^2}{\Nx^2},
\end{equation}
which is the ratio of the high-energy neutrino flux when self-capture is possible to 
the high-energy neutrino flux without the possibility of self-capture.  My primary results 
are illustrations of the dependence of $\beta$ on the parameters $\sigx$, $\sigv$, the 
spin-independent dark matter particle-proton elastic scatting cross section $\sigpsi$, and 
$\Mx$.  To evaluate $\beta$, I {\em do not} assume that the equilibrium solutions of 
Eq.~(\ref{eq:riccati}) are attained.  Rather, I evaluate $\Nx(t=\tau_{\odot})$ and 
$\Nxs(t=\tau_{\odot})$ using the general solution of Eq.~(\ref{eq:rsol2}).

I assume that the cross section for dark matter scattering off of nuclei other than Hydrogen is given 
by \cite{jungman_etal96}
\begin{equation}
\label{eq:sigmasi}
\sign^{\mathrm{SI}} = \sigpsi A^2 \ \frac{\Mx^2 \mn^2}{(\Mx + \mn)^2}\ \frac{(\Mx + \mproton)^2}{\Mx^2 \mproton^2},
\end{equation}
where $A$ is the atomic mass number and $\mproton$ is the proton mass.  Loss of coherence is 
accounted for in the full formulae through suppression by an exponential form factor \cite{gould87a,gould92}.
In the following examples, I focus primarily on spin-independent interactions.  Spin-dependent capture 
of dark matter off of nuclei occurs only for Hydrogen within the Sun and is typically down by roughly two or 
more orders of magnitude at fixed cross section for high-mass ($\Mx \gtrsim 100$~GeV) dark matter particles.  
This may be mitigated by the fact that the spin-dependent cross section for scattering off of protons 
is typically $\sim 1-2$ orders of magnitude larger than the spin-independent cross section in 
viable regions of the constrained minimal 
supersymmetric standard model parameter space \cite{roszkowski_etal07}.  
Moreover, direct search experiments typically use large nuclei with no net spin and exploit the scaling of 
Eq.~(\ref{eq:sigmasi}), so neutrino telescopes \cite{desai_etal04,ackermann_etal06,abbasi_etal09} a very 
competitive with direct search bounds on a spin-dependent interaction 
\cite{lee_etal07,behnke_etal08,angle_etal08,ahmed_etal09} 
and should remain so \cite{wikstrom_edsjo09}.  In \S~\ref{section:results}, I show estimates for the 
spin-independent case as it is more general, including capture off of all nuclei within the Sun, and more 
interesting for present purposes in the sense that 
complementary constraints from direct search experiments are more competitive 
with indirect methods for spin-independent capture.  Including spin-dependent 
capture would typically add a term that is at most comparable 
to $\Cc$ (though this is a model-dependent statement) and I find comparable values of 
$\beta$ for spin-dependent capture with spin-dependent cross sections 
$\sigma_{\mathrm{p}}^{\mathrm{SD}} \sim 10^{2} \sigpsi$.

It is important to set the scale of the signal relative to current and future observations, 
so I present estimates of absolute fluxes in \S~\ref{section:results} as well.  Annihilations in 
the Sun, lead to a flux of high-energy neutrinos at the Earth.  The observable 
signal at a detector such as IceCube \cite{achterberg_etal06,hubert_etal07,abbasi_etal09} is a 
flux of upward-directed high-energy muons induced by scattering of muon neutrinos near the detector.  
The muon flux at the detector is therefore a relatively complicated product and may be written as 
\begin{eqnarray}
\label{eq:phidet}
\Phi & = & \frac{\Gamma_{\mathrm{a}} n_{\mathrm{T}}}{4 \pi A_{\oplus}^2}\ \int_{\Eth}\ \dd E_{\mu}\ \int_{\Eth}\ \dd E_{\nu}\ \nonumber \\
     & \times & \int_{E_{\mu}}^{E_{\nu}}\ \dd \bar{E}_{\mu}\ P(\bar{E}_{\mu} \rightarrow E_{\mu},\lambda)\ 
                 \frac{\dd \sigma_{\nu \mu}(E_{\nu})}{\dd \bar{E}_{\mu}} \nonumber \\
     & \times & \sum_{\mathrm{i}}\ P_{\mathrm{osc}}(\mathrm{i} \rightarrow \mu)\ 
                 \sum_{\mathrm{f}}\ B_{\mathrm{f}}\frac{\dd N_{\mathrm{i/f}}}{\dd E_{\nu}}.
\end{eqnarray}
Individually, the factors in Eq.~(\ref{eq:phidet}) are relatively simple to interpret.  $\Gamma_{\mathrm{a}}$ is 
the annihilation rate in the solar interior, $A_{\oplus}$ is the semi-major axis of the Earth's orbit about the 
Sun, and $n_{\mathrm{T}}$ is the number density of target nuclei near the detector.  The quantity 
$P(\bar{E}_{\mu} \rightarrow E_{\mu},\lambda)$ is the probability for a muon of initial energy $\bar{E}_{\mu}$ 
to have final energy $E_{\mu}$ after traversing a path of length $\lambda$ in the detector material, 
the differential cross section $\dd \sigma_{\nu \mu}(E_{\nu})/\dd \bar{E}_{\mu}$ describes the production 
of a muon of initial energy $\bar{E}_{\mu}$ from an incident neutrino of energy $E_{\nu}$, 
$P_{\mathrm{osc}}(\mathrm{i} \rightarrow \mu)$ is the probability that a neutrino produced as 
flavor $\mathrm{i}$ is a muon neutrino near the detector, $B_{\mathrm{f}}$ is the branching ratio 
to annihilation channel $f$, and $\dd N_{\mathrm{i/f}}/\dd E_{\nu}$ is the differential spectrum of 
neutrinos of flavor $\mathrm{i}$ per unit energy $\dd E_{\nu}$ produced per $\mathrm{f}$-channel annihilation.

I have evaluated Eq.~(\ref{eq:phidet}) for an experiment such as 
IceCube \cite{achterberg_etal06,hubert_etal07,abbasi_etal09} using 
the results of the {\tt WimpSim} Monte Carlo simulations \cite{Blennow_etal08} 
as available through the {\tt DarkSusy} package \cite{gondolo_etal04}.  
I choose $\Eth = 1$~GeV in accord with the common convention for reporting results 
from neutrino telescopes.  An instrument like IceCube observes events above 
tens of GeV, so sensitivities quoted relative to $\Eth = 1$~GeV depend upon 
an assumed spectrum.  I show flux normalizations for two simple choices of 
branching fraction.  I show results for annihilation to $W^+ W^-$ gauge bosons only 
($B_{\mathrm{W^+ W^-}}=1$) as a simple approximation of fluxes that may be produced from annihilation of a 
typical neutralino and a slightly more optimistic case of annihilation 
to $\tau^+ \tau^-$ only ($B_{\mathrm{\tau^+ \tau^-}}=1$).  
Annihilation to $\tau^+ \tau^-$ yields about three times higher flux than annihilation to gauge bosons 
through most of the relevant dark matter particle mass range \cite{Blennow_etal08,wikstrom_edsjo09}.

\section{Results for High-Energy Neutrino Flux Enhancements}
\label{section:results}

\begin{figure*}[t]
\begin{center}
\includegraphics[height=6.7cm]{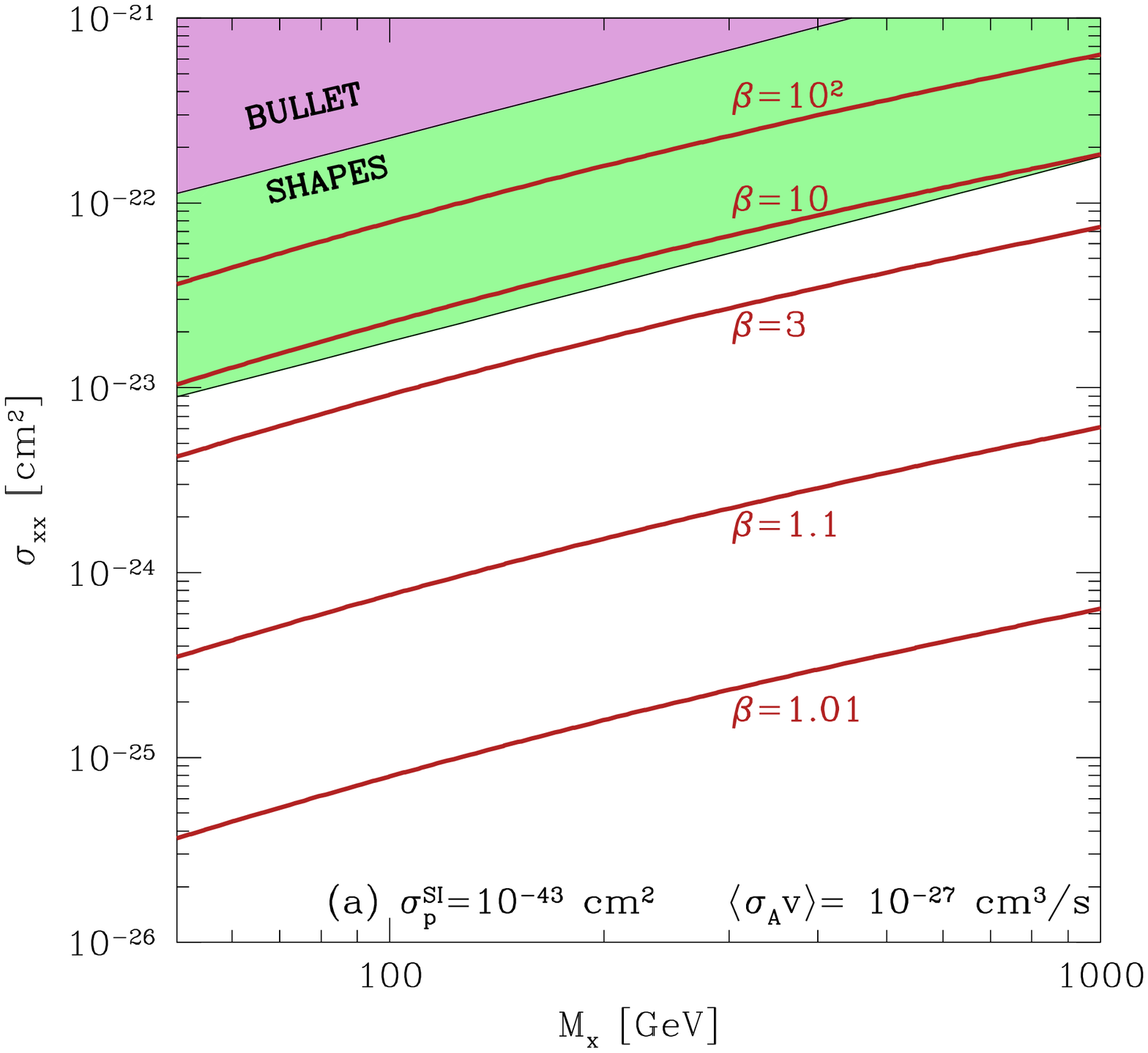}
\vspace*{-20pt}
\includegraphics[height=6.7cm]{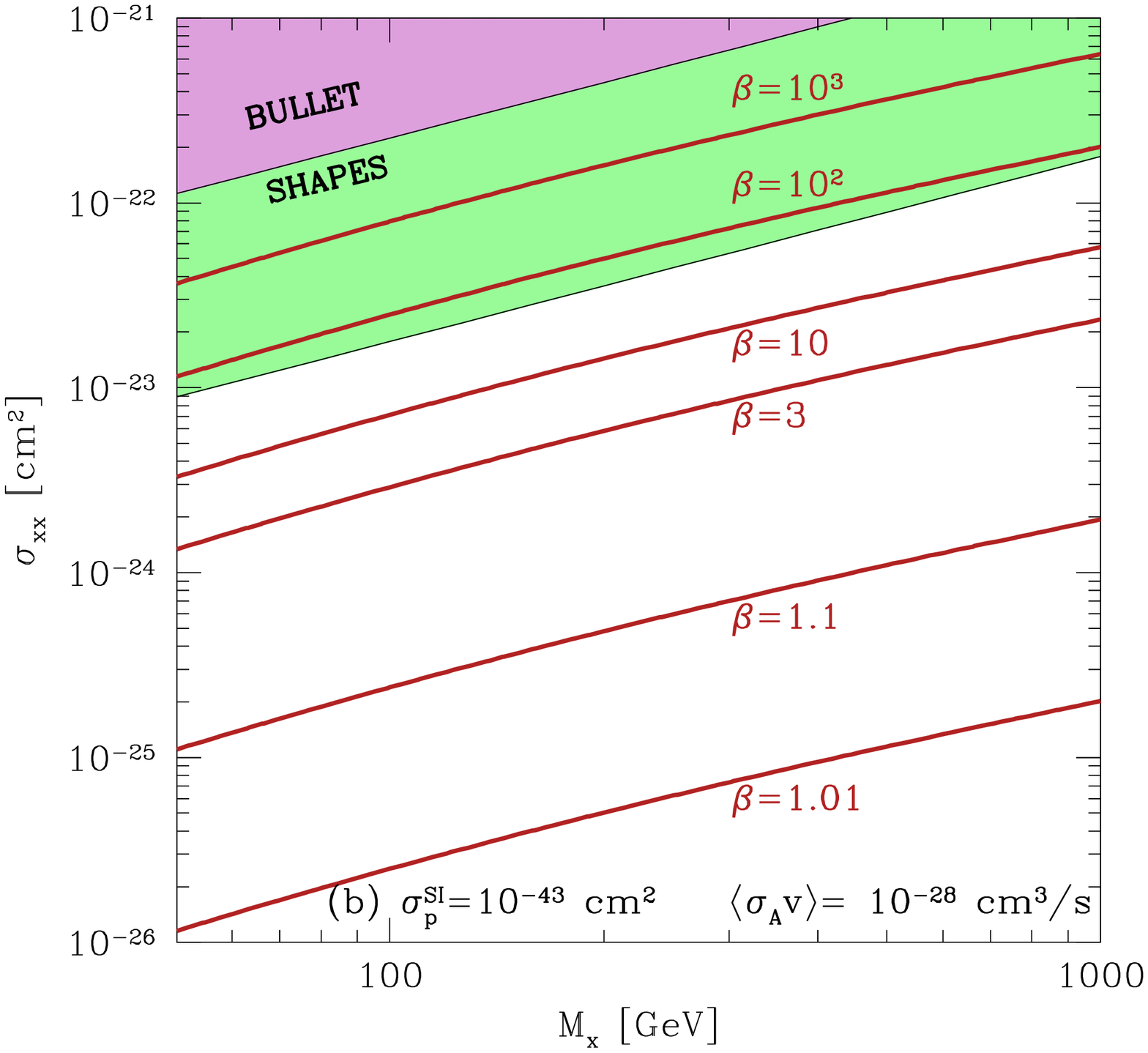}
\vspace*{-20pt}
\includegraphics[height=6.7cm]{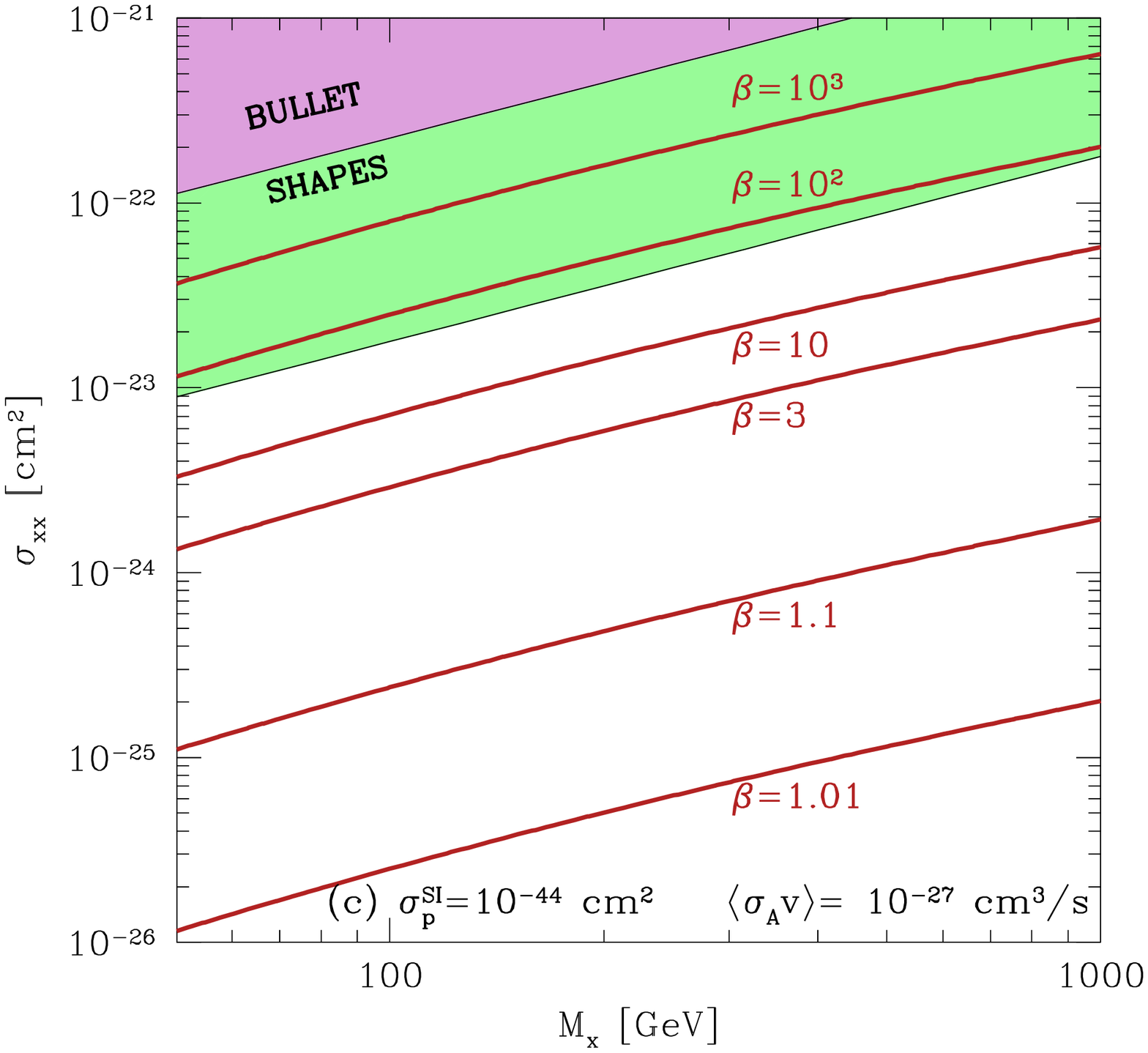}
\includegraphics[height=6.7cm]{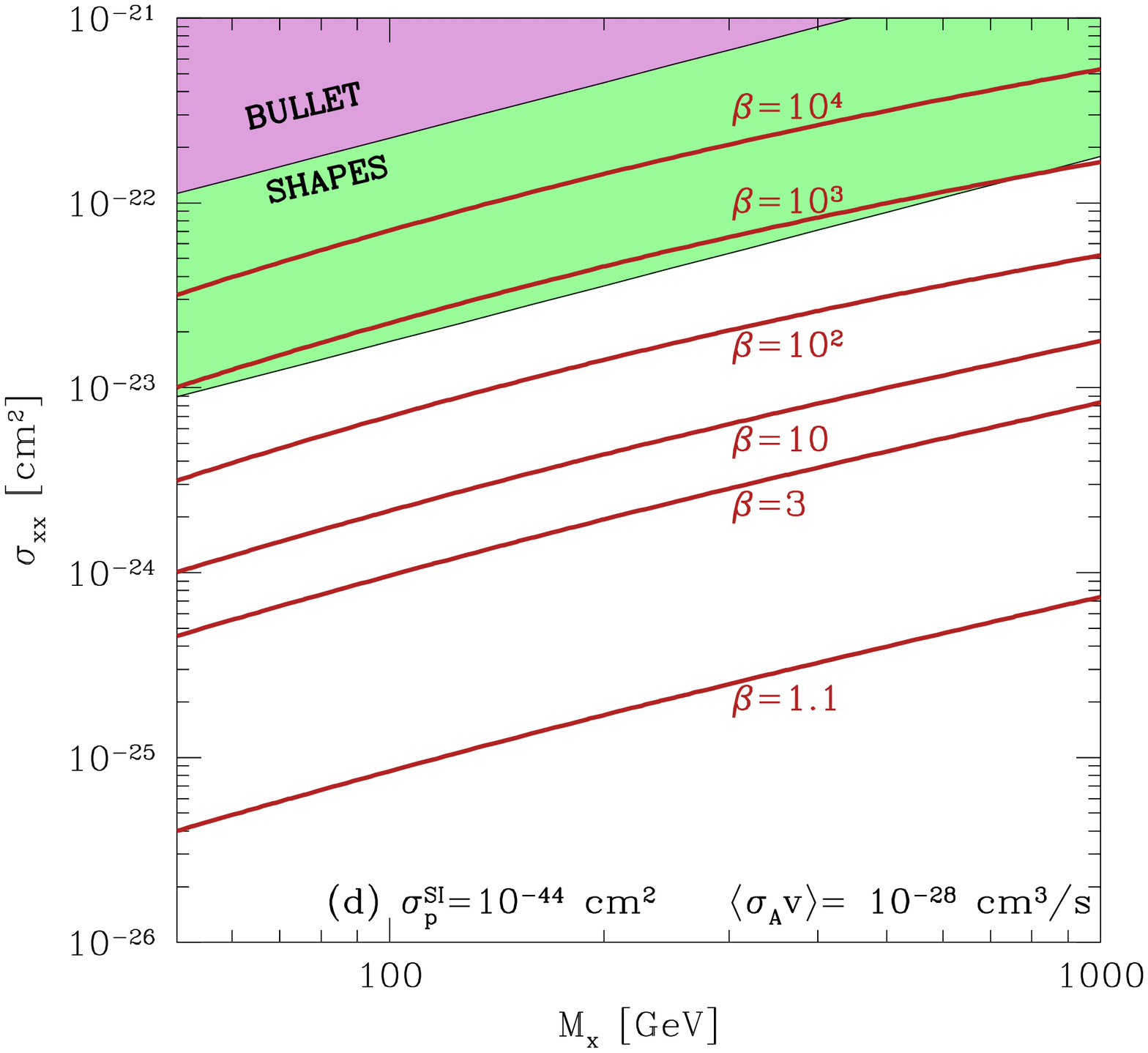}
\includegraphics[height=6.7cm]{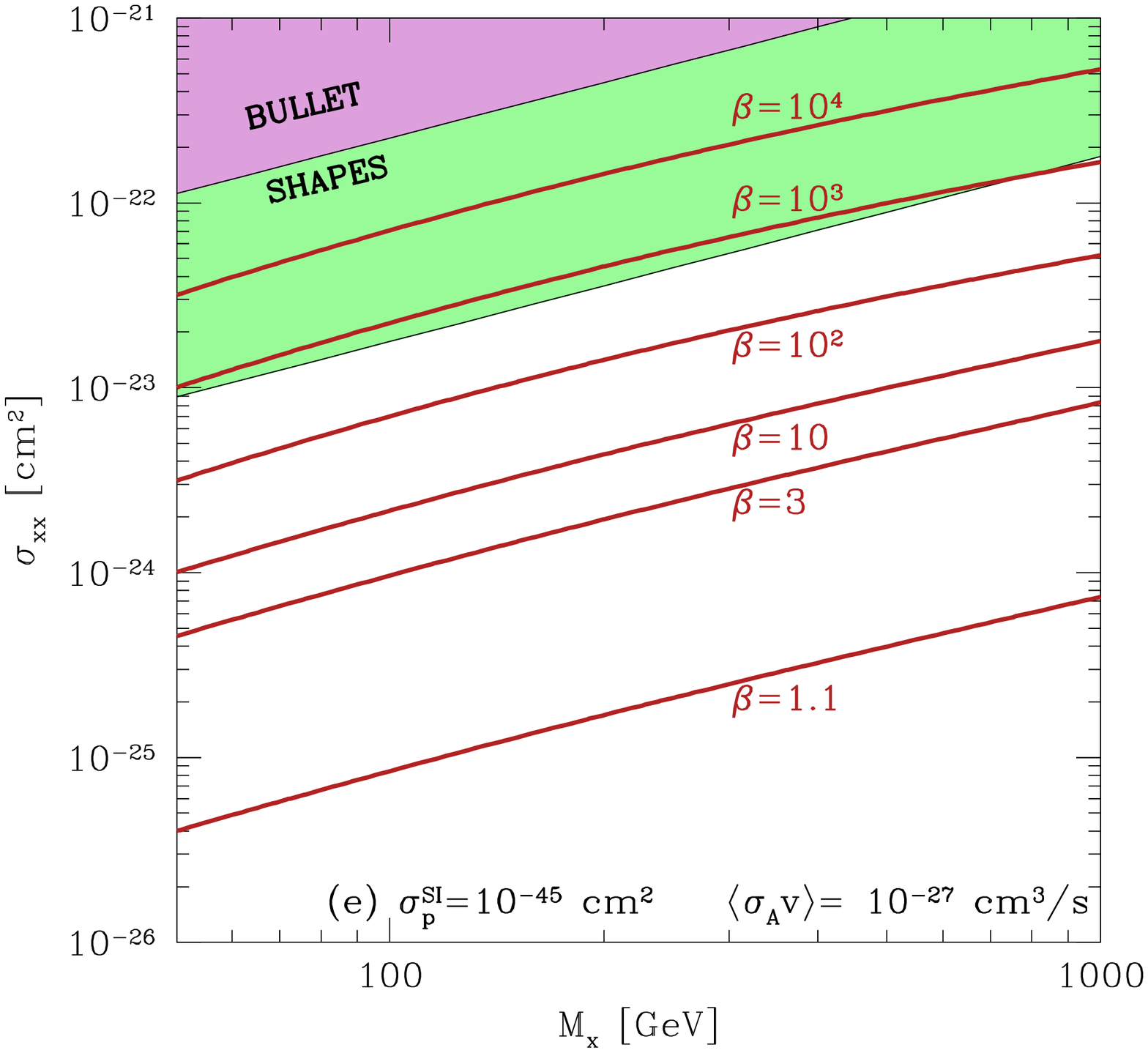}
\includegraphics[height=6.7cm]{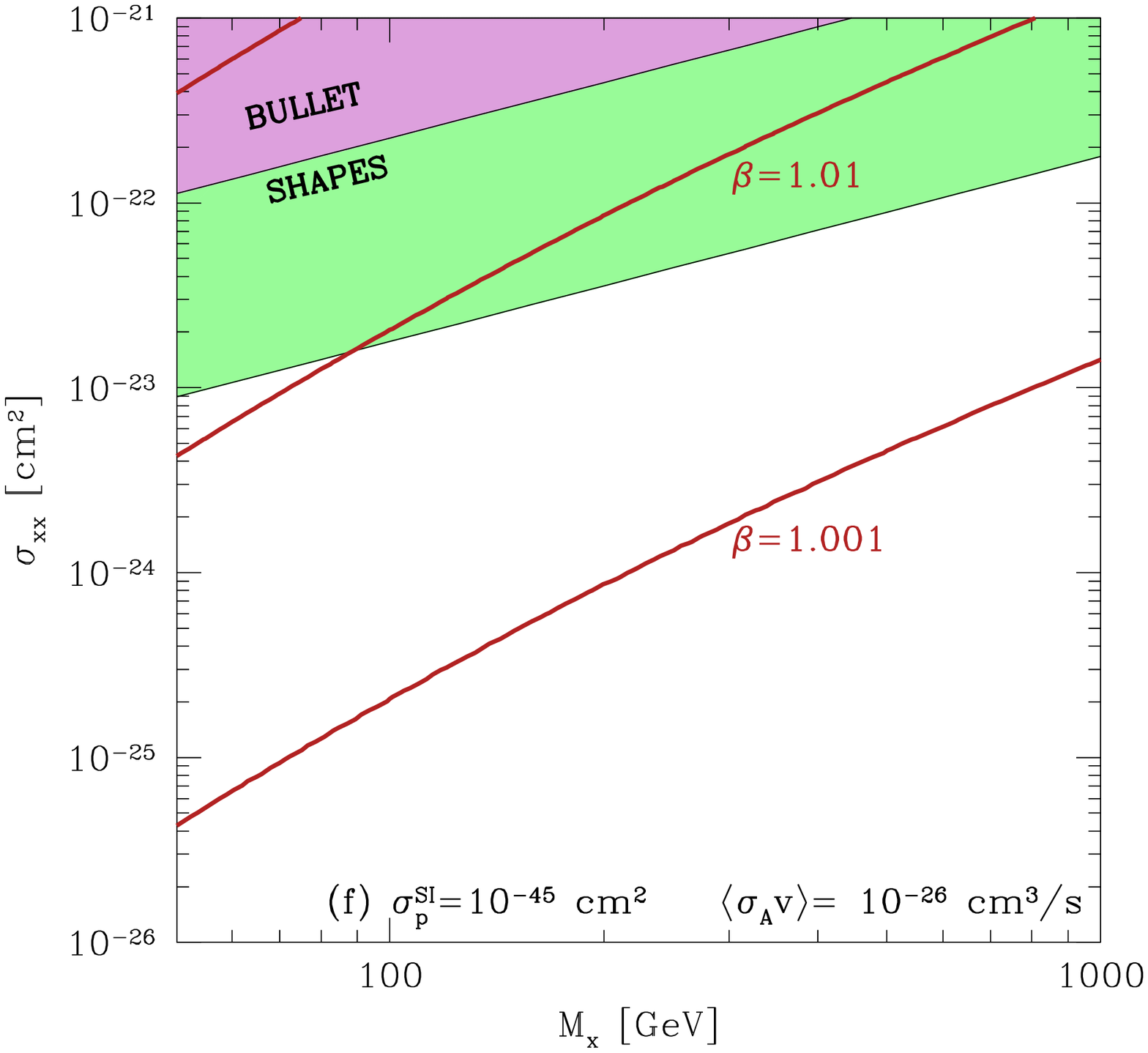}
\caption{
Factors of flux enhancement in the $\Mx$-$\sigx$ plane.  Each panel 
shows contours of constant relative flux enhancement in models of 
self-interacting dark matter.  The panels (a)-(f) are labeled with 
assumed values of $\sigpsi$ and $\sigv$.  The shaded regions at the 
upper left correspond to parameter values that are disfavored by 
analysis of either the Bullet Cluster \cite{randall_etal08} 
or galaxy cluster shapes \cite{yoshida_etal00,dave_etal01,gnedin_ostriker01,miralda-escude02}.
}
\label{fig:sxmx}
\end{center}
\end{figure*}

\begin{figure*}[t]
\begin{center}
\includegraphics[height=7.5cm]{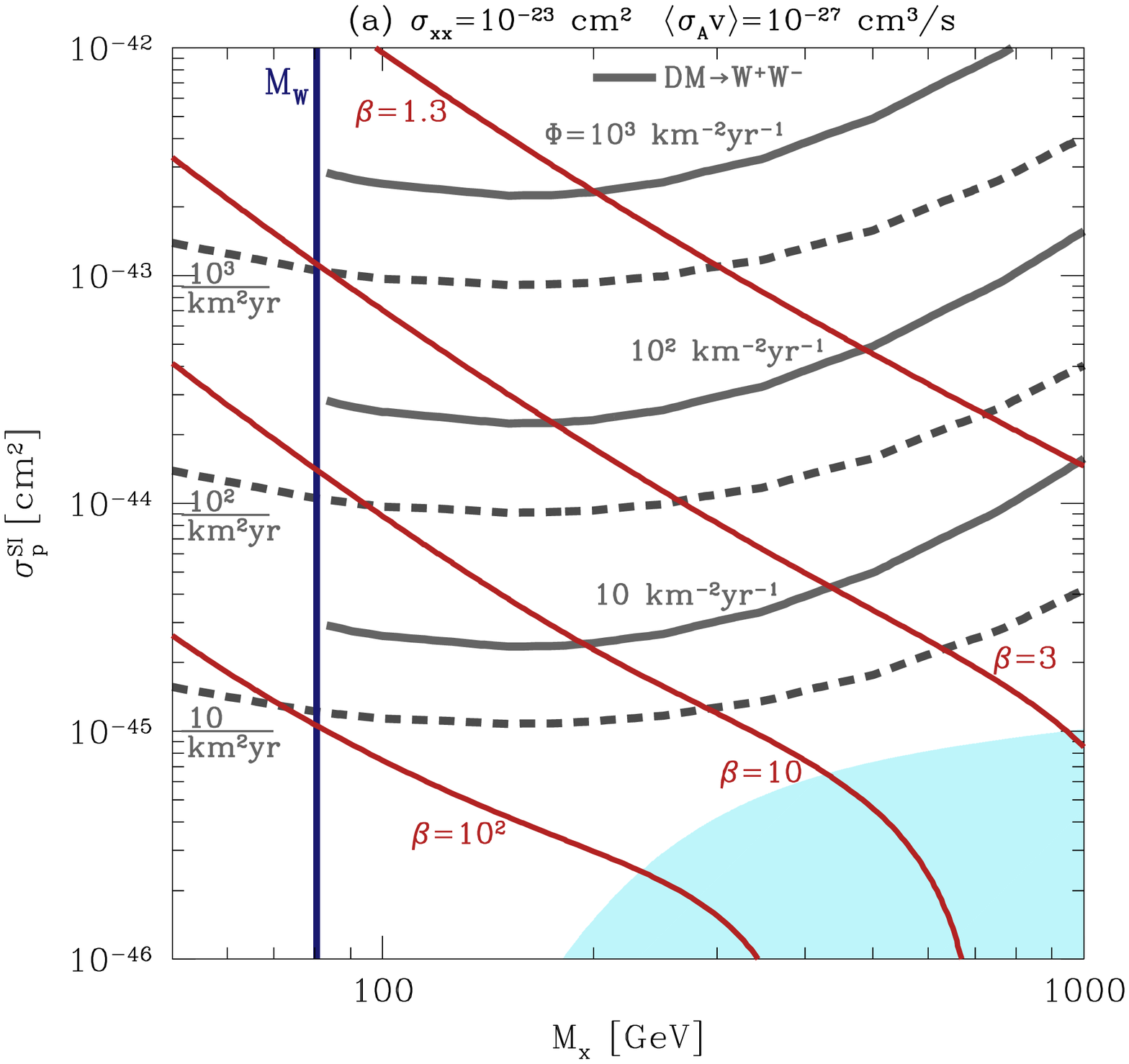}
\vspace*{-8pt}
\includegraphics[height=7.5cm]{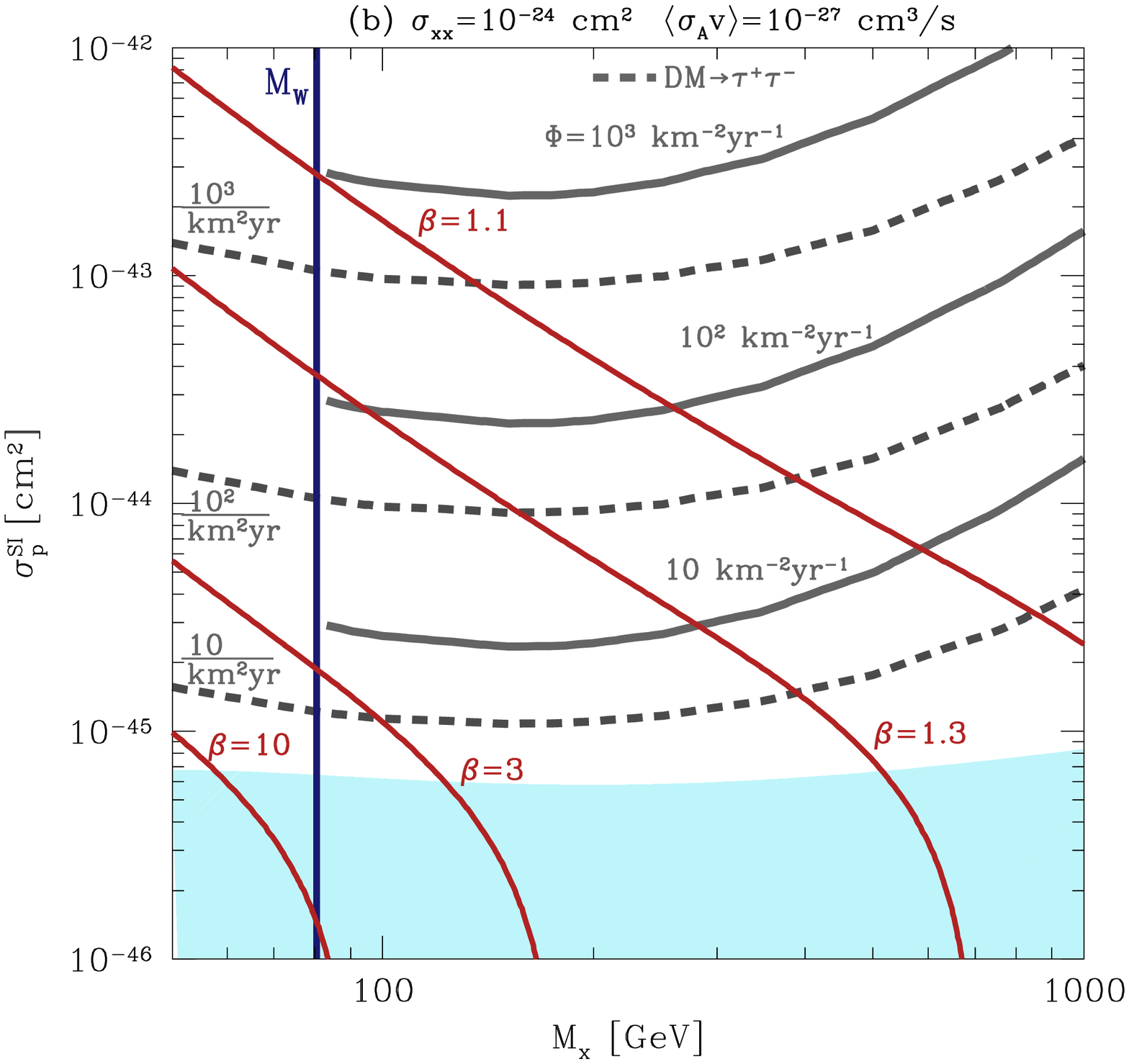}
\includegraphics[height=7.5cm]{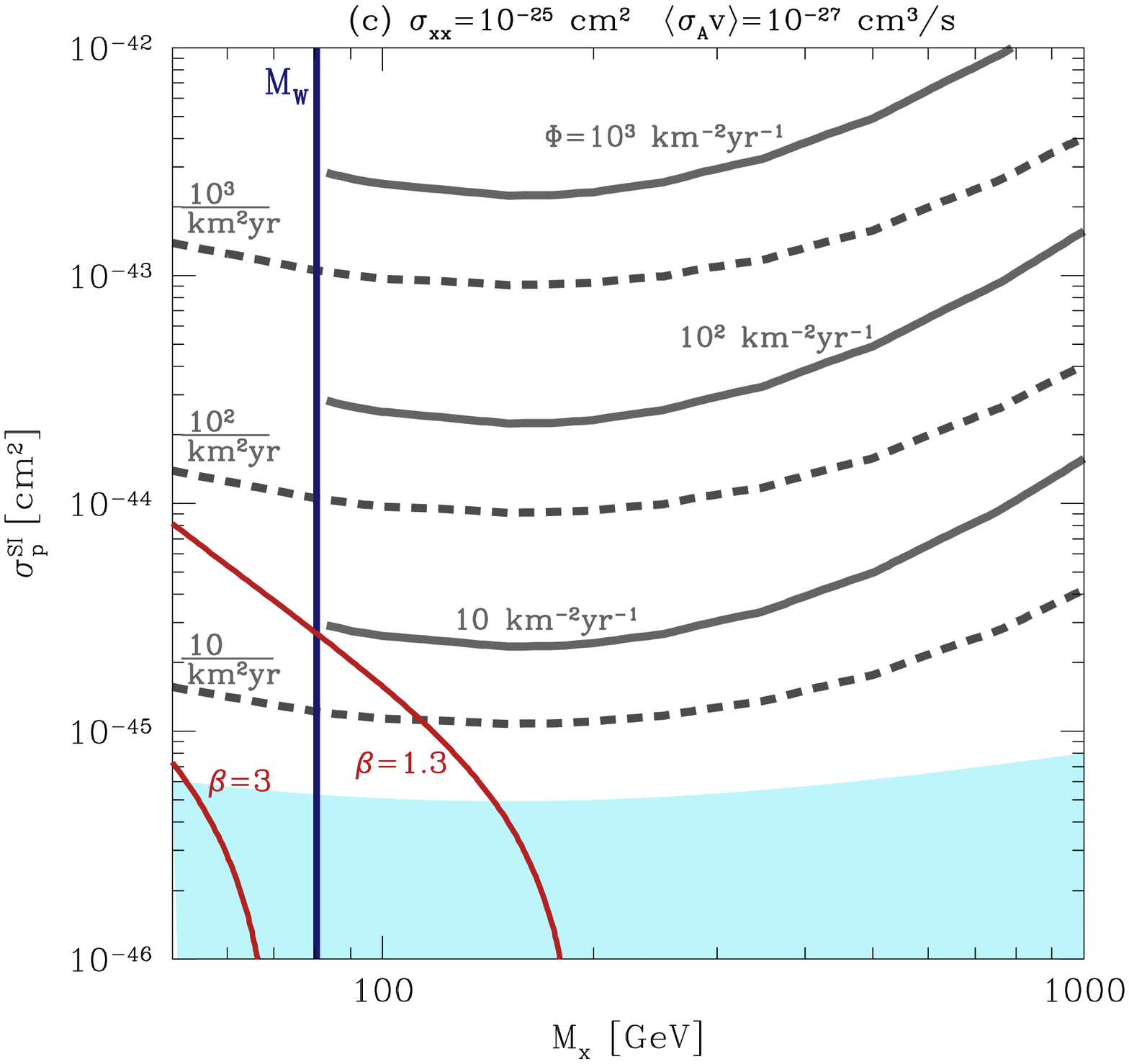}
\includegraphics[height=7.5cm]{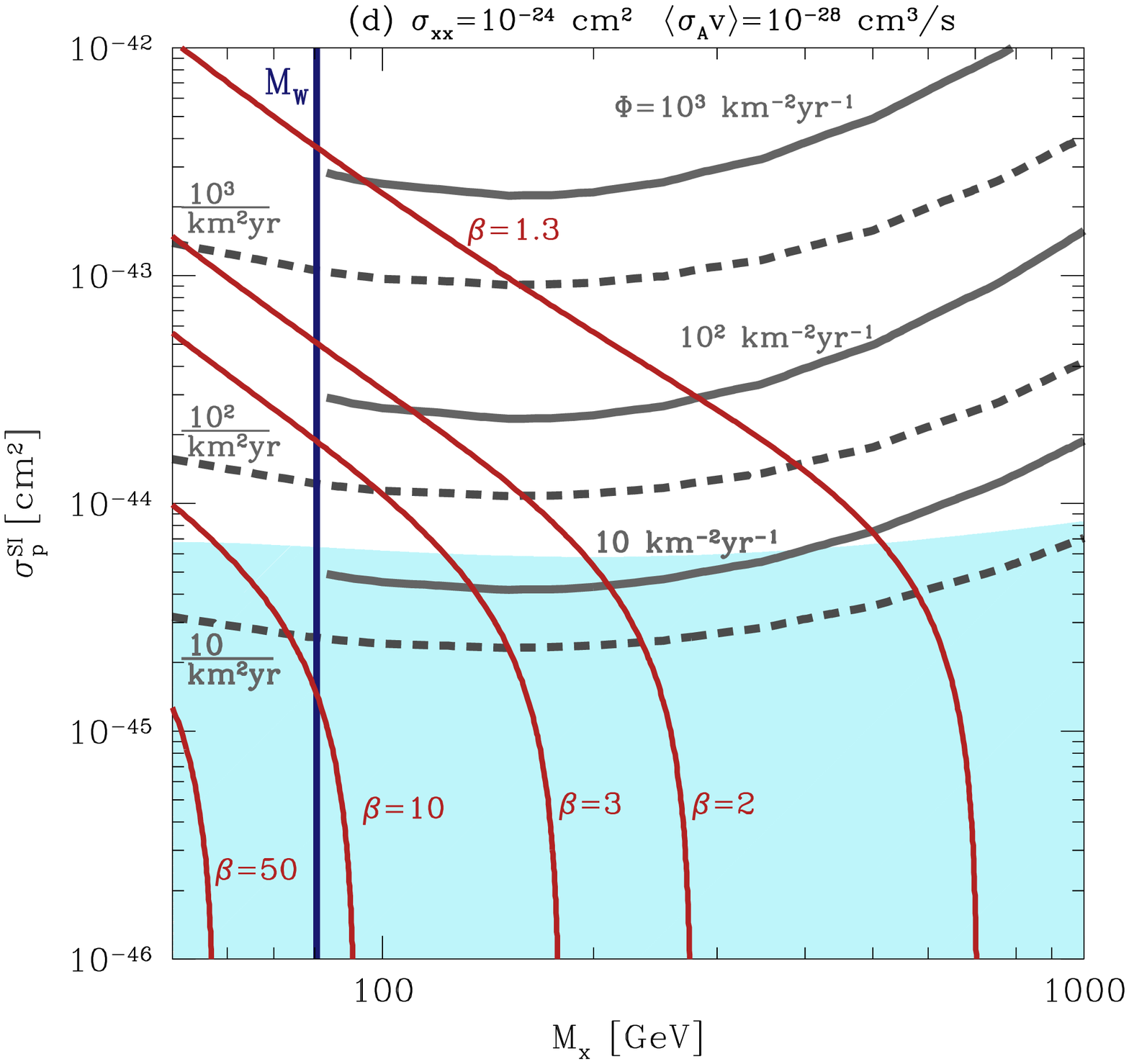}
\caption{
Factors of flux enhancement in the $\Mx$-$\sigpsi$ plane.  Each panel 
shows contours of constant relative flux enhancement in models of 
self-interacting dark matter.  The flux enhancement contours 
are the {\em red lines with negative slope} labeled by the flux 
enhancement factor $\beta$.  Every panel is labeled according to 
the assumed values of $\sigx$ and $\sigv$ in each calculation.  
I show for reference on the background in each plot contours of 
constant muon flux at a detector on Earth in the case where 
annihilation happens through $W^+W^-$ ({\em solid, with cut-off 
at the $W$ mass)} and the case where annihilation happens 
through $\tau^+\tau^-$ ({\em dashed}).  These reference flux 
levels are computed assuming that there is no significant 
self-capture of dark matter, $\Cs=0$.  Shaded regions at the 
lower ends of these plots correspond to models that are 
not yet at their equilibrium 
levels for a Sun of age $\tau_{\odot}=5 \times 10^9$~yr.  
Roughly speaking, current direct dark matter 
searches constrain the spin-independent dark matter-proton cross section 
to slightly better than $\sigpsi \gtrsim 10^{-43} \cmsq$ 
at about $10^2$~GeV \cite{angle_etal08,ahmed_etal09}.  
}
\label{fig:spmx}
\end{center}
\end{figure*}

I summarize results on the relative importance of a contribution from dark matter particle self-capture 
to high-energy neutrino fluxes from the Sun in the 
contour plots of Figure~\ref{fig:sxmx} and Figure~\ref{fig:spmx}.  
There are four parameters of most immediate interest, namely $\sigx$, $\Mx$, $\sigpsi$, and $\sigv$.  
Fig.~\ref{fig:sxmx} displays contours of constant $\beta$ in the $\Mx$-$\sigx$ plane for several 
fixed values of $\sigpsi$ and $\sigv$, while Fig.~\ref{fig:spmx} shows contours of $\beta$ in the 
$\Mx$-$\sigpsi$ plane for specific choices of $\sigx$ and $\sigv$.

Consider Fig.~\ref{fig:sxmx}, which shows an interesting, general set of results.  
The shaded regions at the upper left represent values of $\sigx$ that 
are already ruled out \cite{yoshida_etal00,dave_etal01,gnedin_ostriker01,miralda-escude02,randall_etal08}; 
however, computing these bounds is complex and the position of the boundary in each case 
remains somewhat controversial \cite{yoshida_etal00,gnedin_ostriker01,randall_etal08} so this boundary 
should be regarded as approximate.  Notice that boosts in neutrino fluxes of 
several tens up to 100\% can be achieved for quite reasonable 
parameter values.  Significantly larger boosts of up to $\beta \sim 10^3$ can be realized in 
extreme regions of the viable parameter space.

Beyond this, several additional features of Fig.~\ref{fig:sxmx} are worthy of explicit note.  
Eq.~(\ref{eq:rsnum}) indicates that lines of constant $\beta$ on the $\Mx$-$\sigx$ plane should 
run as $\sigx \propto \Mx^{5/4}$.  In practice, the lines of constant $\beta$ are somewhat more shallow 
than this because the approximate form of $\Cc$ [Eq.~(\ref{eq:Cselftext})] 
used in the simple estimate of Eq.~(\ref{eq:rsnum}) assumed favorable 
kinematics for scattering off of nuclei at all dark matter particle masses.  
In the case of favorable kinematics, $\Cc \propto \Mx^{-1}$.  However, as the 
dark matter particle and nucleus masses become less comparable, the capture rate 
tends to $\Cc \propto \Mx^{-2}$ and lines of constant $\beta$ become shallower, 
approaching $\sigx \propto \Mx^{3/4}$.  In addition, comparing the pair of panels 
(b) and (c) or (d) and (e) in Fig.~\ref{fig:sxmx} it is clear that the scaling 
$\sigx^2 \propto \sign \sigv$ from Eq.~(\ref{eq:rsnum}) 
for fixed $\beta$ at a particular $\Mx$ is valid.  This is sensible, because 
these cross sections serve only to scale the rates $\Cs$, $\Cc$, and $\Ca$ 
(so long as the equilibrium solution is achieved).

Also evident in Fig.~\ref{fig:sxmx} is that large enhancements may only be achieved when 
annihilation cross sections are relatively low $\sigv < 10^{-26}\ \mathrm{cm}^3\mathrm{s}^{-1}$, 
where the numerical value is the canonical value for a thermal relic dark matter particle.  
This can be seen most dramatically in panel (f) where I have taken 
$\sigpsi=10^{-45}\ \mathrm{cm}^2$ and $\sigv=10^{-26}\ \mathrm{cm}^3\mathrm{s}^{-1}$ and 
the enhancements are at most a few percent over viable parameter ranges.  Even 
discounting experimental limitations, it is thought that the intrinsic errors in computing 
neutrino fluxes from dark matter capture within the Sun should be a few tens of percent 
\cite{gould87a,gould92,duda_etal07,peter_tremaine08,bruch_etal09,peter09b}, 
so this indicates that such an effect can only be interesting when 
$\sigv \lesssim \mathrm{a}\ \mathrm{few}\ \times 10^{-27} \mathrm{cm}^3\mathrm{s}^{-1}$.  
Within the context of scans of restricted regions of supersymmetric parameter space, 
annihilation cross sections well below this value are achievable \cite{baltz_gondolo04,
roszkowski_etal07,baltz_etal08,martinez_etal09}, so significantly lower cross sections 
are attainable in theories with complex particle spectra.  Even in exceedingly simple 
proposals there exists sufficient freedom to set $\sigx$ and $\sigv$ 
apart significantly \cite{spergel_steinhardt00,arkani-hamed_etal08a}.  Consider interaction 
via exchange of a boson of mass $\mboson$.  The perturbative annihilation and scattering 
cross sections should be related as $\siga/\sigx \sim (\mboson/\Mx)^4$ and both 
$\mboson$ and the coupling strength remain to be fixed.  However, these results do indicate that models 
of dark matter self-interaction that lead to very large annihilation cross sections (such 
as the Sommerfeld-enhanced scenarios of significant recent interest, see 
Refs.~\cite{hisano_etal04,hisano_etal05,arkani-hamed_etal08a,lattanzi_silk08}) will 
induce little additional neutrino flux due to self-capture.

Figure~\ref{fig:spmx} shows contours of $\beta$ in the $\Mx$-$\sigpsi$ plane for four choices of 
$\sigx$ and $\sigv$ and complements the results in Fig.~\ref{fig:sxmx}.  First, the results of 
current direct dark matter searches can be compared in this plane.  Direct search experiments constrain 
$\sigpsi$ directly.  Current bounds place $\sigpsi \gtrsim 10^{-43} \cmsq$ at $\Mx \sim 10^2$~GeV 
\cite{lee_etal07,angle_etal08,behnke_etal08,ahmed_etal09}.  This bound becomes slightly 
better with decreasing $\Mx$ until $\Mx \sim 50$~GeV and at higher masses this bound grows $\propto \Mx$.  
The competition is much less severe from spin-dependent searches \cite{lee_etal07,behnke_etal08,ahmed_etal09}, 
as the indirect limit from the Sun already exceeds the direct search 
limit by more than an order of magnitude over a wide range of masses \cite{abbasi_etal09}.  
Second, Fig.~\ref{fig:spmx} shows contours of constant 
absolute muon flux above a threshold of $\Eth=1$~GeV in models where 
there is no self-interaction in order to set the absolute scale.  Two annihilation channels are shown, 
annihilation into $W^+W^-$ (solid gray) and $\tau^+\tau^-$ (dashed gray).  Current experiments 
are limited to muon fluxes above several hundred per km$^2$ per year 
\cite{achterberg_etal06,hubert_etal07,abbasi_etal09}.  Assuming the relatively hard 
spectra from annihilation to gauge bosons, IceCube with DeepCore extension 
should optimistically be capable of detecting fluxes down to $\sim 60$~km$^{-2}$yr$^{-1}$ 
for particle masses above $\sim 200$~GeV, with relatively lower sensitivity below this mass 
\cite{cowen05,achterberg_etal06,hubert_etal07,cowen09,wikstrom_edsjo09}.  
A deep-sea neutrino facility such as the {\tt KM3NeT} 
effort \cite{carr_etal08}, building on the {\tt ANTARES} 
\cite{montaruli09,montaruli09b,lahmann09,ageron09,carminati09}, 
{\tt NEMO} \cite{capone03,riccobene_etal04,riccobene06}, and {\tt NESTOR} 
\cite{aggouras_etal06b,aggouras_etal06} work, 
may achieve comparable or better sensitivities.

Figure~\ref{fig:spmx} also shows regions where the equilibrium solution of Eq.~(\ref{eq:rsteady2}) 
has not yet been attained for a solar age of $\tau_{\odot}=5 \times 10^9$~Gyr.  I approximate 
the equilibrium boundary as the contour where the predicted flux is 58\% of the value it would 
be at equilibrium, because $\tanh^2(1) \simeq 0.58$ [see Eq.~(\ref{eq:rsteady1}).  
Equilibrium is achieved in the majority of the parameter 
space corresponding to a potentially-detectable signal \cite{wikstrom_edsjo09}.  
However, notice that the contours of constant flux at Earth in scenarios 
with no self-capture are not the same in each panel [particularly so in panel (d)] because 
fluxes are no longer determined solely by $\sigpsi$ and $\Mx$ for models that are not 
equilibrated.

Consider panels, (b)-(d) of Fig.~\ref{fig:spmx}.  In these panels, 
the equilibrium boundary exhibits a very shallow minimum in $\sigpsi$ 
near $\Mx$ of a few hundred GeV.  In these panels, the equilibrium boundaries 
are essentially the same as they would be in the absence of any dark matter self interaction.  
The minimum occurs due to the competition between capture and annihilation in the relevant 
timescale, $\tau_{\mathrm{eq}} = 1/\sqrt{\Cc \Ca}$.  The annihilation rate scales with dark matter 
mass as $\Ca \propto \Mx^{3/2}$ [see Eq.~(\ref{eq:ca2}) and Eq.~(\ref{eq:veff})], while 
at relatively low masses $\Cc \propto \Mx^{-1}$.  For dark matter particle masses greater than 
several hundred GeV, the capture rate transitions to the regime where kinematic suppression of 
capture becomes important and $\Cc \propto \Mx^{-2}$ (I have neglected the orbital effects that 
also tend to slow thermalization within the Sun for high-mass dark matter candidates \cite{peter_tremaine08}).

On the other hand, panel (a) of Fig.~\ref{fig:spmx} shows a distinct feature in the equilibration boundary at 
$\Mx \sim 300$~GeV.  The feature is caused by self-capture.  At high $\sigx$ and low $\Mx$, self-capture 
of dark matter is important and can drive rapid equilibration even for very low values of $\Cc$.  
In the absence of self-capture, the equilibration boundary in panel (a) of Fig.~\ref{fig:spmx} would 
be relatively flat as a function of $\Mx$ as in panels (b)-(d).

The contours of constant $\beta$ in Fig.~\ref{fig:spmx} show that interesting regions of parameter 
space can lead to detectable boosts in muon fluxes at Earth of tens of percent to 100\%.  Somewhat 
more extreme choices of parameters can lead to boosts of an order of magnitude or more.  In particular 
regions of the parameter space, the dark matter self-interaction can drive a model that would be 
undetectable or ruled out by direct searches in terms of $\sigpsi$ to be detectable at contemporary 
or future high-energy neutrino telescopes.  This is an interesting possibility, because 
this implies that the indirect neutrino signal from the Sun would not be related to either direct 
search results or indirect neutrino signals form the Earth in a straightforward manner.  
Each contour of $\beta$ in Fig.~\ref{fig:spmx} exhibits a distinct 
break as it nears the equilibration boundary.  This is because solutions 
that are well away from equilibrium tend to lie in the linear portion of $\Nx(t)$, prior to 
any significant opportunity for exponential growth (see Fig.~\ref{fig:nuex}).  As a result, the 
insight gained from Eq.~(\ref{eq:rsnum}) fails at low cross-sections and large enhancement factors 
require significantly smaller $\sigpsi$ at fixed $\Mx$ than one would estimate from the equilibrium 
assumption.

\begin{figure}[t]
\includegraphics[height=8.0cm]{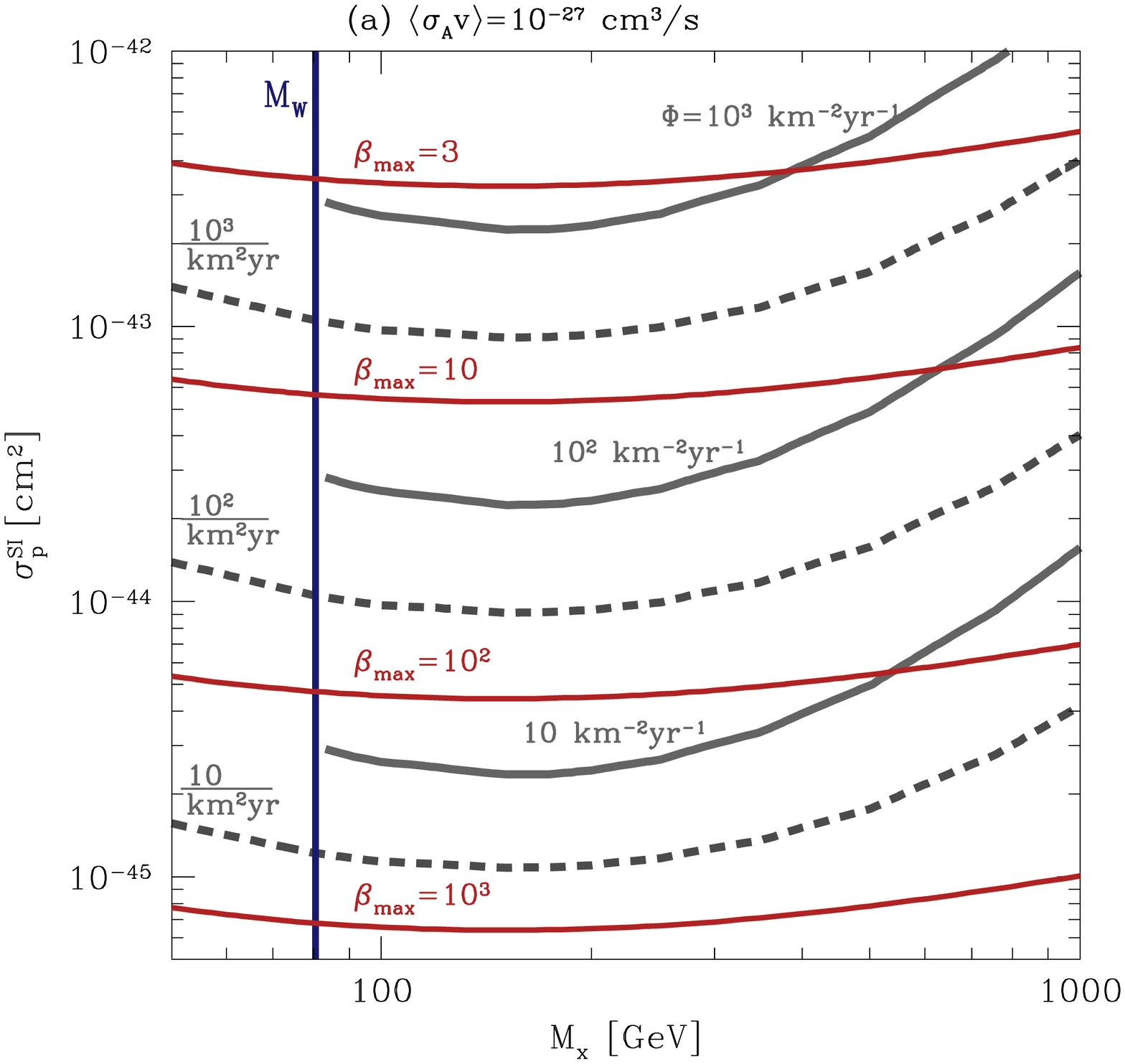}
\caption{
Contours of maximum possible flux enhancements given 
existing bounds on dark matter elastic scattering cross 
sections.  This figure is similar to the figure panels of 
Fig.~\ref{fig:spmx}.  However, in this figure, I show contours 
of $\beta_{\mathrm{max}}$, the maximum possible flux enhancement 
at each point in the $\Mx-\sigpsi$ plane.  I compute this maximal 
boost at each point by setting $\sigx$ to the maximum allowed value 
at each value of $\Mx$ (see the limits in Fig.~\ref{fig:sxmx}).  
The quantity $\beta_{\mathrm{max}}$ is also a function of 
annihilation cross section and this panel shows $\beta_{\mathrm{max}}$ 
for $\sigv = 10^{-27}\ \mathrm{cm}^3\mathrm{s}^{-1}$.  
From this figure, it is already clear that certain combinations 
of parameters may be ruled out with contemporary or forthcoming 
neutrino telescope data (e.g., Ref.~\cite{abbasi_etal09}).
}
\label{fig:spmax}
\end{figure}

Of course, it is likely that the effect of flux enhancement due to self-capture is 
negligible; however, it is useful to know just how large this effect 
could possibly be.  It is simple to make such an estimate and contours 
of the maximum possible flux enhancement $\beta_{\mathrm{max}}$, are shown 
in Figure~\ref{fig:spmax}.  The ``maximum possible flux enhancement,'' 
depends upon $\sigv$ and I compute it as follows.  
At each value of $\Mx$, I choose the largest 
value of $\sigx$ that is not already excluded by considerations of 
large-scale structure (see the contours in Fig.~\ref{fig:sxmx}).  
I then compute the flux enhancements at each point in the 
$\Mx$-$\sigpsi$ plane for a fixed $\sigv$ 
($\sigv=10^{-27}\ \mathrm{cm}^3\mathrm{s}^{-1}$ in this case).  
The enhancement scales approximately as $\sim \sigv^{-1}$ 
as given in Eq.~(\ref{eq:rssym}).

Fig.~\ref{fig:spmax} already illustrates 
that some extreme parameter combinations may be ruled out with 
contemporary or near future limits from neutrino telescopes.
Notice also that the contours of constant $\beta_{\mathrm{max}}$ are 
very flat functions of $\Mx$.  This is because existing limits 
on dark matter particle self-interactions scale as $\sigx \propto \Mx$.  
This is important for comparison with direct detection experiments 
which aim to achieve limits on the dark matter-proton scattering cross 
section on the order of $\sigpsi \sim 10^{-44} \cmsq$ in the near 
future \cite{lee_etal07,angle_etal08,behnke_etal08,ahmed_etal09}.  
Absent dark matter self-capture, such a limit would indicate that 
there should be no observable high-energy neutrino flux from the 
Sun, but self-capture can clearly modify this conclusion.

\section{Summary and Discussion}
\label{section:discussion}

In this paper I have reconsidered the indirect high-energy neutrino signal from within 
the Sun in models in which the dark matter particles have significant self interactions.  
The influence of self-interactions is that they may allow dark matter particles within the 
Galactic halo to be captured within the Sun by scattering off of dark matter particles that have 
already been captured by scattering off of nuclei within the Sun.  For sufficiently large 
dark matter self-interaction cross sections, this can lead to a period during which the rate 
of capture of dark matter particles by the Sun grows in proportion to the number of dark 
matter particles already captured by the Sun.  The number of dark matter particles within the 
Sun then grows exponentially until this increase is stopped by efficient annihilation.  
The net result is that the Sun may contain significantly more dark matter than in models with 
no dark matter self interaction and high-energy neutrino signals due to annihilation of 
these particles may be significantly higher as a result.

In \S~\ref{section:results}, I showed that mild enhancements of a few tens to one hundred percent 
are possible over a wide range of viable parameter space that may be probed with 
contemporary and near-future neutrino telescopes 
\cite{achterberg_etal06,hubert_etal07,montaruli09,riccobene_etal04,riccobene06,aggouras_etal06b}.  
Significantly larger flux enhancements of up to a factor of $\sim 10^2$ are possible in more 
extreme corners of the dark matter parameter space and at flux levels that are not yet 
within reach of near-term neutrino telescopes.  Ten percent enhancements are not 
particularly interesting at present because intrinsic errors in the flux predictions 
are at the tens of percent level \cite{gould87a,gould92,duda_etal07,peter_tremaine08,peter09b}, 
but this situation may improve as experimental advancements drive renewed interest in this signal.

Large flux enhancements require large dark matter self-interaction cross 
sections $\sigx \gtrsim 10^{-24} \cmsq$ and relatively small dark matter mutual annihilation 
cross sections $\sigv \lesssim 10^{-27}\ \mathrm{cm}^3\mathrm{s}^{-1}$.  The small annihilation 
cross section allows the number of dark matter particles within the Sun to grow exponentially 
for a prolonged period of time, which enables large flux enhancements.  More specifically, the 
neutrino flux enhancement grows approximately as $\propto \sigx^2/\sigpsi \sigv$, 
neglecting the possibility that for some values of these parameters the flux may not reach its 
equilibrium level for a sun of age $\tau_{\odot} = 5 \times 10^9$~yr.  As a consequence, 
large enhancements require a disparity between scattering and annihilation cross 
sections that may be unfamiliar.  However, such a disparity is practicable and, in fact, 
previous proposals of self-interacting dark matter rely on just such relative differences 
in cross sections in order to produce significant astrophysical effects without annihilating 
all of the dark matter in the early universe (e.g., Refs.~\cite{spergel_steinhardt00,arkani-hamed_etal08a}).

The high-energy neutrino flux enhancement I compute may have several interesting implications.  
In \S~\ref{section:sc} and in the Appendix, I show that the flux from 
within the Earth will not be enhanced due to dark matter self interactions.  
In the standard picture, the flux from within the Earth can be predicted relative to the 
solar flux.  In viable contemporary models, the Earth signal is often not yet 
equilibrated and the Sun-to-Earth flux ratio depends upon the dark matter 
particle mass as well as the capture and annihilation rates.  
In the self-interacting scenario, the relation between the Sun and Earth neutrino 
fluxes may no longer hold and significant deviations from any predicted ratio 
may be a sign of dark matter self interactions.

Likewise, experiments that undertake direct dark matter searches may exploit indirect 
detection methods to cross-check limits and/or detections.  The correspondence between 
direct detection experiments and high-energy neutrinos from the Sun is relatively straightforward.  
Though direct detection rates and high-energy neutrino fluxes depend on somewhat different 
integrals over the dark matter velocity distribution, in the standard picture they 
both grow in proportion to the product of the local dark matter density multiplied by 
the dark matter-nucleon cross section, $\propto \rhox \sign$.  If dark matter exhibits 
considerable self-interaction, this correspondence is also broken.  The neutrino flux 
from the Sun may be significantly larger than would be predicted based on the limits or 
detections from direct detection experiments.  One extreme possibility is that neutrino fluxes 
that may seemingly be ruled out by direct searches (based upon limits on $\sigpsi$ and/or $\sigpsd$) 
may be realized due to the enhancement from dark matter self-interactions.  A broken 
correspondence between the neutrino fluxes and direct search results may signal dark matter 
interactions.

The pace of the quest to identify the dark matter is picking up rapidly.  Neutrino 
telescopes play an important role in this endeavor and the indirect limits from 
existing facilities are already competitive with direct search techniques.  
The indirect, high-energy neutrino signal from the 
Sun may serve as a unique probe of new physics confined to the dark sector, 
and experimental advancements in the near future should shed new light on the 
properties of the dark matter.



\appendix*
\section{Capture and Self-Capture of Dark Matter Particles in the Sun}

In the interest of completeness, I give a brief discussion of the capture of self-interacting 
dark matter particles within the Sun in this appendix.  The treatment here is not original, 
save for the fact that I consider dark matter particles interacting among themselves, 
and follows the lucid discussion given in the series of papers by A. Gould 
\cite{gould87a,gould87b,gould92}.  I conclude this section with the rate of 
dark matter particle self-capture.  For the results in the main text, 
I use the full formulae of Ref.~\cite{gould92} 
to compute dark matter particle capture off of nucleons.

Gould begins by considering capture in an individual spherical shell of the body 
on which capture is occurring (the Sun in this case) of radius $r$ and local 
escape speed $\vescr$.  About this shell, consider a bounding surface of radius $R$ 
so large that the gravitational field due to the Sun is negligible at $R$.  
Let the one-dimensional speed distribution function of dark matter particles at 
this shell be $f(u)$, where $u$ is the speed at infinity and the 
integral of $f(u)$ over all speeds gives the number density of dark matter particles.  
The inward flux of particles of speed $u$ at angle 
$\theta$ relative to radial across the surface at $R$ is then
\begin{equation}
\label{eq:fin}
\frac{\dd {\mathcal{F}}_{\mathrm{in}}}{\dd u \ \dd \cos^2(\theta)} = \frac{1}{4} f(u) u.
\end{equation}
Changing variables from $\cos^2(\theta)$ to the specific angular momentum 
$J=R u \sin(\theta)$, and integrating over the surface area of the sphere at 
$R$ gives the rate at which dark matter particles enter the surface per 
unit time, per unit speed, per unit angular momentum, 
\begin{equation}
\label{eq:rin}
\frac{\dd {\mathcal{R}}_{\mathrm{in}}}{\dd u \ \dd J^2} = \frac{\pi f(u)}{ u}.
\end{equation}
Notice that I have written this so that the quantity 
${\mathcal{R}}_{\mathrm{in}}$ has dimensions of inverse time.  

Take $\Omega(w)$ to be the rate at which a particle with speed $w$ at the shell at $r$ scatters 
to a speed less than $\vescr$.  Infalling dark matter particles with 
speed at $R$ of $u$ that reach the shell at $r$, do so with speed 
\begin{equation}
\label{eq:wshell}
w = \sqrt{u^2 + \vesc^2(r)}.
\end{equation}
The probability of such a particle to be captured is 
\begin{equation}
\label{eq:pcapsh}
\dd P = \frac{\Omega(w)}{w} \frac{2 \dd r}{\sqrt{1-\frac{J^2}{r^2 w^2}}} \Theta(rw-J),
\end{equation}
where the quantity $2 \dd r$ multiplied by the term under the radical is the path length through 
the shell, dividing by $w$ converts this to the time spent in the shell, $\Theta(x)$ is a 
step function and the particular step function above enforces the condition that only 
particles with $J < rw$ intersect the shell.  Multiplying the rate of incoming particles in 
Eq.~(\ref{eq:rin}) with Eq.~(\ref{eq:pcapsh}), the differential rate of capture within 
the shell is 
\begin{eqnarray}
\label{eq:dCdrdJ}
\frac{\dd C}{\dd r \ \dd u \ \dd J^2} & = & \frac{\dd \mathcal{R}_{\mathrm{in}}}{\dd u\ \dd J^2}\ \frac{\dd P}{\dd r}  \\
     & = & \frac{2 \pi f(u)}{ w u } \frac{\Omega(w)} 
{\sqrt{1 - \frac{J^2}{r^2 w^2}}}\ \Theta ( rw - J ). \nonumber
\end{eqnarray}
The integral over $J^2$ can be performed leaving the capture rate per unit speed at 
infinity, per unit shell volume
\begin{equation}
\label{eq:dCdrdu}
\frac{\dd C}{\dd u \dd V} = \frac{f(u)}{u} w \Omega(w),
\end{equation}
where I have replaced $ 4\pi r^2 \dd r$ with $\dd V$.  This gives the rate per unit shell 
volume as an integral over the speed distribution at infinity,  
\begin{equation}
\label{eq:dCdV}
\frac{\dd C}{\dd V} = \int\ \frac{f(u)}{u} w \Omega(w)\ \dd u,
\end{equation}
and the task remains to determine $\Omega(w)$, perform the integration 
over speeds in Eq.~(\ref{eq:dCdV}), and integrate over the volume of the Sun.

The rate of scattering in the shell is simply $n \sigma w$, with $\sigma$ the scattering 
cross section and $n$ the number density of targets.  The case of most practical interest 
is velocity-independent and nearly isotropic scattering of infalling dark matter particles 
against targets that are effectively at rest with respect to the capturing body.  
In this case, the fractional loss of kinetic energy in a 
given scattering event is a uniform distribution over the interval 
\begin{equation}
\label{eq:fracenergy}
0 \le \frac{\Delta E}{E} \le \frac{4 \Mx m}{(\Mx + m)^2},
\end{equation}
where $\Mx$ is the mass of the dark matter particle and $m$ is the mass of the particle it 
scatters off of.  The dark matter particle must lose a fraction of its kinetic 
energy $\Delta E/E > u^2 / w^2$ in order to be captured.  If the condition
\begin{equation}
\label{eq:cond}
4 \Mx m/(\Mx +m)^2 \ge u^2 /w^2 
\end{equation}
holds, the probability that an individual 
scattering event leads to capture is 
\begin{equation}
\label{eq:pcap}
p_{\mathrm{cap}} = \frac{\vesc^2(r)}{w^2}\Bigg[1-\frac{u^2}{\vesc^2(r)} \frac{(\Mx -m)^2}{4 \Mx m}\Bigg].
\end{equation}
Therefore, if Eq.~(\ref{eq:cond}) holds, 
\begin{equation}
\label{eq:Omega}
\Omega(w)=n \sigma \vescr \frac{\vescr}{w}\Bigg[1-\frac{u^2}{\vesc^2(r)} \frac{(\Mx-m)^2}{4\Mx m}\Bigg].
\end{equation}
At least one property of Eq.~(\ref{eq:Omega}) is familiar.   Capture is most efficient when both 
projectile and target are of the same mass and becomes less efficient as the masses become mismatched.

Combining Eq.~(\ref{eq:dCdV}) with Eq.~(\ref{eq:Omega}) yields the capture rate per shell volume 
in the Sun, 
\begin{equation}
\label{eq:dCdV2}
\frac{\dd C}{\dd V} = \int \ n \sigma \vesc^2(r) \frac{f(u)}{u} 
\Bigg[1 - \frac{u^2}{\vesc^2(r)} \frac{(\Mx-m)^2}{4\Mx m} \Bigg] \dd u.
\end{equation}
Gould has evaluated this expression for the case of a Maxwell-Boltzmann speed 
distribution including possible form-factor suppression of scattering with large nuclei at 
high momentum transfer \cite{gould87a,gould92}.  However, the general formulae are rather unwieldy, the integrations are 
lengthy but straightforward, and presenting them does not add significantly to the 
insight needed for my purposes.  As a result, I will not present the general formulae and 
will move to a particularly simple special case.

Of particular interest for the present paper is the capture of dark matter particles in the halo by 
other dark matter particles that have already been captured within the Sun.  As a consequence, I will evaluate 
Eq.~(\ref{eq:dCdV2}) for the special case of $m=\Mx$ and for capture by the Sun moving 
with speed $\vsun=220 \kms$ through a Maxwell-Boltzmann distribution of dark matter particles 
with dispersion $\vbar=270 \kms$.  The distribution function can then be written 
\begin{equation}
\label{eq:fumb}
f(x)= \frac{2 \nx}{\sqrt{\pi}} x^2 e^{-x^2} e^{-\eta^2} \frac{\sinh( 2 x \eta )}{ x \eta},
\end{equation}
in terms of the dimensionless variables $x^2 = 3 (u/\vbar)^2/2$ and $\eta^2 = 3 (\vsun/\vbar)^2/2$.
Integrating over the speed distribution yields 
\begin{equation}
\label{eq:dCdVxx}
\frac{\dd C}{\dd V} = \sqrt{\frac{3}{2}} \nx \ n \sigma \vescr \ \frac{\vescr}{\vbar} \frac{\erf(\eta)}{\eta}.
\end{equation}

The total capture rate now requires integrating over the volume of the Sun.  This gives  
\begin{eqnarray}
\label{eq:totintegral}
C & = & \sqrt{\frac{3}{2}}\ \nx \sigma \vesc(\Rsun)\ \frac{\vesc(\Rsun)}{\vbar} \frac{\erf(\eta)}{\eta} \nonumber \\
  & \times & \int_0^{\Rsun} 4 \pi r^2 n \frac{\vesc^2(r)}{\vesc^2(\Rsun)} \dd r.
\end{eqnarray}
The last integral can be re-written conveniently by defining a 
dimensionless potential $\hat{\phi}=\vesc^2(r)/\vesc^2(\Rsun)$, 
in which case the last integral is the product of the total number of targets $N$ and the average of 
$\hat{\phi}$ over all targets within the Sun, 
\begin{equation}
\label{eq:captot}
C = \sqrt{\frac{3}{2}}\ \nx \sigma \vesc(\Rsun)\ \frac{\vesc(\Rsun)}{\vbar} N \langle \hat{\phi} \rangle 
\frac{\erf(\eta)}{\eta}.
\end{equation}
The numerical factor of $\sqrt{3/2}$ in Eq.~(\ref{eq:captot}) differs from the factor 
$\sqrt{6/\pi}$ given in Refs.~\cite{gould87a,gould87b,gould92} because Gould defined 
the error function $\erf(x)$ with an unconventional normalization.

I have assumed that $\Mx=m$ to derive Eq.~(\ref{eq:captot}).  However, so long as the mass of the target and 
and projectile are not very mismatched, scattering will be likely to lead to capture and 
Eq.~(\ref{eq:captot}) will be a relatively good approximation for the capture rate.  
In the case of capture by scattering off of nuclei, the relevant cross section is the 
elastic scattering cross section off of the nucleus of interest $\sigma=\sign$ and $N$ is the number of such nuclei 
in the Sun.  The total capture rate due to scattering off of all nuclei is the sum of the individual rates for 
all of the different nuclear species within the Sun.

For dark matter self-capture, the relevant cross section is the elastic scattering cross section of dark matter 
particles with themselves $\sigma=\sigx$ and $N=\Nx$ is the number of dark matter particles already captured within the 
Sun.  Therefore, the dark matter self-capture rate coefficient referred to in the main text can be approximated as 
\begin{equation}
\label{eq:Cselfcoeff}
\Cs = \sqrt{\frac{3}{2}}\ \nx \sigx \vesc(\Rsun)\ \frac{\vesc(\Rsun)}{\vbar} 
\langle \hat{\phi}_{\mathrm{x}} \rangle \frac{\erf(\eta)}{\eta}.
\end{equation}
As discussed in the text, captured dark matter particles typically occupy a very small range of radii 
within the Sun (typically confined to only a few percent of $\Rsun$), in which case 
$\langle \hat{\phi}_{\mathrm{x}} \rangle \simeq 5.1$ \cite{gould92}.

In the case of dark matter particle self-capture, there is one additional complication that 
must be accounted for that is not relevant for capture off of nuclei.  The Sun is optically-thin 
to the propagation of dark matter particles, so a target dark matter particle that receives 
too much kinetic energy relative to the solar core will be ejected resulting in no net gain 
of dark matter particles.  Therefore, not only must the collision result in an energy 
exchange of $\Delta E/E \ge u^2/w^2$, but it must be limited to $\Delta E/E \le \vesc^2(r)/w^2$.  
This modifies the capture probability per collision (again, taking $m=\Mx$) to 
\begin{equation}
\label{eq:pcapfull}
p_{\mathrm{cap}} = \Bigg( \frac{\vesc^2(r) - u^2}{w^2} \Bigg)\ \Theta( \vesc(r) - u)
\end{equation}
and the capture rate to 
\begin{equation}
\label{eq:ratefull}
\Omega(w)=\frac{n\sigma}{w} (\vesc^2(r)-u^2).
\end{equation}
When $\vesc(r) \gg u$, this modification is relatively minor.  This is because in this situation, 
the incoming dark matter particle must only lose a small fraction of its total energy to be captured 
and does not necessarily impart enough energy to escape on the target dark matter particle.  This 
is generally the case for the Sun, because escape from the solar interior requires speeds at least 
two times larger than the typical speed at infinity of a dark matter particle.  However, the escape 
speed from the Earth is significantly smaller than the typical speeds of dark matter particles, 
so collisions within the Earth that lead to capture of the infalling particle will almost 
always lead to ejection of the target.  In fact, most interactions of this kind will lead 
to both infalling particle and target being unbound from the Earth.  An 
interesting question is to ask whether self-interactions may 
scour the Earth of captured dark matter particles, but a comparison of 
the relevant rates along the lines leading to Eq.~(\ref{eq:rsnum}) in \S~\ref{section:sc} 
shows that the removal rate is significantly less than the capture rate for parameters 
of interest.

\begin{widetext}

This small modification results in a significantly more complex formula for the rate of capture.  The 
calculation follows according to the simple estimate given above.  Again, the integrations are 
lengthy but straightforward, so I will only quote the result.  The full rate of capture accounting 
for the potential recoil and ejection of the target dark matter particles is 
\begin{eqnarray}
\label{eq:Cselffinal}
\Cs & = & \sqrt{\frac{3}{2}}\ \nx \sigx \vesc(\Rsun) \frac{\vesc(\Rsun)}{\vbar} \eta^{-1} \nonumber \\
    & \times & \Bigg( \Bigg[ \langle \hat{\phi}_{\mathrm{x}}\rangle \erf(\eta) 
    - \frac{(\langle \hat{\phi}_{\mathrm{x}} \erf(\xv+\eta) \rangle 
    - \langle \hat{\phi}_{\mathrm{x}} \erf(\xv-\eta)\rangle)}{2} \Bigg] \nonumber \\
    &   & - \frac{2}{3\sqrt{\pi}}\Bigg(\frac{\vbar}{\vesc(\Rsun)}\Bigg)^2 \eta 
          \Bigg\{ \frac{\sqrt{\pi}}{2}\eta \Bigg( 2 \erf(\eta) - [\langle \erf(\xv+\eta)\rangle - \langle \erf(\xv-\eta)\rangle]\Bigg) \nonumber \\
    &   & + 2e^{-\eta^2} - [\langle e^{(\xv-\eta)^2} \rangle - \langle e^{(\xv + \eta)^2} \rangle ]  
          + \frac{2}{\eta}\mathcal{J}(0,\eta) - \frac{1}{\eta} \langle \mathcal{J}(\xv-\eta,\xv+\eta)\rangle \Bigg\}  \Bigg),
\end{eqnarray}
where $\xv^2 = 3(\vesc(r)/\vbar)^2/2$, the brackets about a quantity, such as ``$\langle q \rangle$,'' 
designate the average over all captured dark matter particles of the quantity $q$, and the integral 
$\mathcal{J}(s,t)=\int_s^t\ q^2 e^{-q^2}\ \dd q$.  The first term in this relation is the simple result 
from Eq.~(\ref{eq:Cselfcoeff}).  The second term in the first set of square braces results 
from truncating the integral over the speed distribution at $u=\vesc(r)$.  Typically, $\xv \pm \eta > 1$, so 
this term will be small in comparison to the first term. The terms within the curly braces come from 
the new piece in the capture rate Eq.~(\ref{eq:ratefull}).  The factor that multiplies the terms in curly 
braces is typically of order $\sim 0.06$ for the Sun.  Consequently, the new terms in Eq~(\ref{eq:Cselffinal}) 
collectively represent relatively small modifications to Eq.~(\ref{eq:Cselfcoeff}).  This fits the heuristic 
understanding that ejection due to recoil will be important only when $ \vesc(r) \lesssim \vbar \sim \vsun$.  
\end{widetext}

\begin{figure}[t]
\begin{center}
\includegraphics[height=7.5cm]{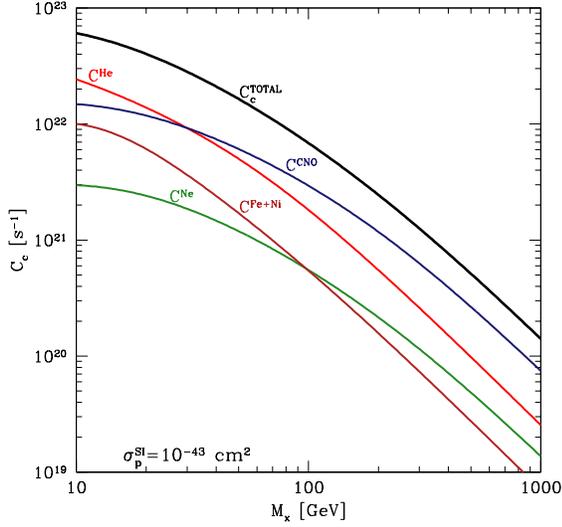}
\caption{
Capture rates of weakly-interacting dark matter particles used in the calculations 
in the main text.  In this panel, I assume $\sigpsi=10^{-43} \cmsq$ and I show capture 
rates as a function of dark matter particle mass.  In addition to the total capture 
rate, $\Cc^{\mathrm{TOTAL}}$, I also show capture rates off of several elements within 
the Sun for those elements most important to capture via a scalar interaction.  These 
are He ($C^{\mathrm{He}}$), the sum of C, O, and N ($C^{\mathrm{CNO}}$, Oxygen is the 
most important of the CNO elements individually), the sum of Fe and Ni ($C^{\mathrm{Fe}+\mathrm{Ni}}$), 
and Ne ($C^{\mathrm{Ne}}$).  At high-mass, the capture rate approaches $\Cc \propto \Mx^{-2}$ 
as expected \cite{gould92}.
}
\label{fig:captrate}
\end{center}
\end{figure}

Though the above sketch of Gould's derivations is instructive for present purposes, the formulae I present here 
do not suffice to make an adequate estimate of capture by nuclei within the Sun.  
In all of the detailed results in \S~\ref{section:results}, I use the full formulae given 
in Ref.~\cite{gould92} and repeated in the review of Ref.~\cite{jungman_etal96}.  I take 
$\vsun=220 \kms$, $\vbar=270 \kms$, $\rhox=0.4 \ \mathrm{GeV}/\mathrm{cm}^3$ \cite{klypin_etal02}, 
the solar mass distribution of Ref.~\cite{gould92}, and 
the elemental abundances given in the review of Ref.~\cite{grevesse_etal07}.  I show a specific 
example of my calculations of the rate of capture of 
dark matter particles from spin-independent scattering off of nuclei in the Sun with a spin-independent 
cross section for dark matter-proton scattering of $\sigpsi=10^{-43} \cmsq$ in Figure~\ref{fig:captrate}.  
In addition to the total capture rate, I show also in Fig.~\ref{fig:captrate} contributions to the total 
capture rate from scattering off of several of the most important nuclei within the Sun.  Capture off of 
Hydrogen is down by roughly two orders of magnitude throughout most of this range due to the lower cross 
section relative to heavier nuclei and the unfavorable scattering kinematics for heavy dark matter particles.

\begin{acknowledgments}

I am thankful to Gianfranco Bertone, Katherine Freese, Dan Hooper, Savvas Koushiappas, 
Brant Robertson, Joe Silk, Louis Strigari, and Tim Whatley for helpful discussions 
and email exchanges.  I am particularly grateful to 
John Beacom and Dan Boyanovsky for a number of detailed and helpful 
discussions regarding an early draft of this manuscript.  This work was supported by the 
University of Pittsburgh, by the National Science Foundation through grant 
AST 0806367, and by the Department of Energy.

\end{acknowledgments}

\bibliography{ms}

\end{document}